\documentclass[acmtog]{acmart}
\acmSubmissionID{405}

\usepackage{booktabs} 

\usepackage[ruled]{algorithm2e} 
\usepackage{amsmath}
\usepackage{xcolor}
\usepackage{makecell}
\usepackage{booktabs}
\usepackage{array}

\SetAlFnt{\small}
\SetAlCapFnt{\small}
\SetAlCapNameFnt{\small}
\SetAlCapHSkip{0pt}
\newcolumntype{P}[1]{>{\centering\arraybackslash}p{#1}}

\copyrightyear{2024}
\acmYear{2024}
\setcopyright{acmlicensed}\acmConference[SA Conference Papers '24]{SIGGRAPH Asia 2024 Conference Papers}{December 3--6, 2024}{Tokyo, Japan}
\acmBooktitle{SIGGRAPH Asia 2024 Conference Papers (SA Conference Papers '24), December 3--6, 2024, Tokyo, Japan}
\acmDOI{10.1145/3680528.3687600}
\acmISBN{979-8-4007-1131-2/24/12}

\begin{document}
\title{Large \'Etendue 3D Holographic Display with Content-adaptive Dynamic Fourier Modulation}

\author{Brian Chao}
\orcid{0000-0002-4581-6850}
\affiliation{
  \institution{Stanford University}
  \city{Stanford}
  \state{CA}
  \postcode{94305}
  \country{USA}}
\email{brianchc@stanford.edu}

\author{Manu Gopakumar}
\orcid{0000-0001-9017-4968}
\affiliation{
  \institution{Stanford University}
  \city{Stanford}
  \state{CA}
  \postcode{94305}
  \country{USA}}
\email{manugopa@stanford.edu}

\author{Suyeon Choi}
\orcid{0000-0001-9030-0960}
\affiliation{
  \institution{Stanford University}
  \city{Stanford}
  \state{CA}
  \postcode{94305}
  \country{USA}}
\email{suyeon@stanford.edu}

\author{Jonghyun Kim}
\orcid{0000-0002-1197-368X}
\affiliation{
  \institution{NVIDIA}
  \city{Santa Clara}
  \state{CA}
  \postcode{95051}
  \country{USA}}
\email{jonghyunk@nvidia.com}

\author{Liang Shi}
\orcid{0000-0002-4442-4679}
\affiliation{
  \institution{Massachusetts Institute of Technology}
  \city{Cambridge}
  \state{MA}
  \postcode{02139}
  \country{USA}}
\email{liangs@mit.edu}

\author{Gordon Wetzstein}
\orcid{0000-0002-9243-6885}
\affiliation{
  \institution{Stanford University}
  \city{Stanford}
  \state{CA}
  \postcode{94305}
  \country{USA}}
\email{gordon.wetzstein@stanford.edu}

\begin{abstract}
Emerging holographic display technology offers unique capabilities for next-generation virtual reality systems. Current holographic near-eye displays, however, only support a small \'etendue, which results in a direct tradeoff between achievable field of view and eyebox size. \'Etendue expansion has recently been explored, but existing approaches are either fundamentally limited in the image quality that can be achieved or they require extremely high-speed spatial light modulators. We describe a new \'etendue expansion approach that combines multiple coherent sources with content-adaptive amplitude modulation of the hologram spectrum in the Fourier plane. To generate time-multiplexed phase and amplitude patterns for our spatial light modulators, we devise a pupil-aware gradient-descent-based computer-generated holography algorithm that is supervised by a large-baseline target light field. Compared with relevant baseline approaches, ours demonstrates significant improvements in image quality and \'etendue in simulation and with an experimental holographic display prototype.
\end{abstract}

%
%


%
%


\begin{teaserfigure}
  \centering
	\includegraphics[width=\columnwidth]{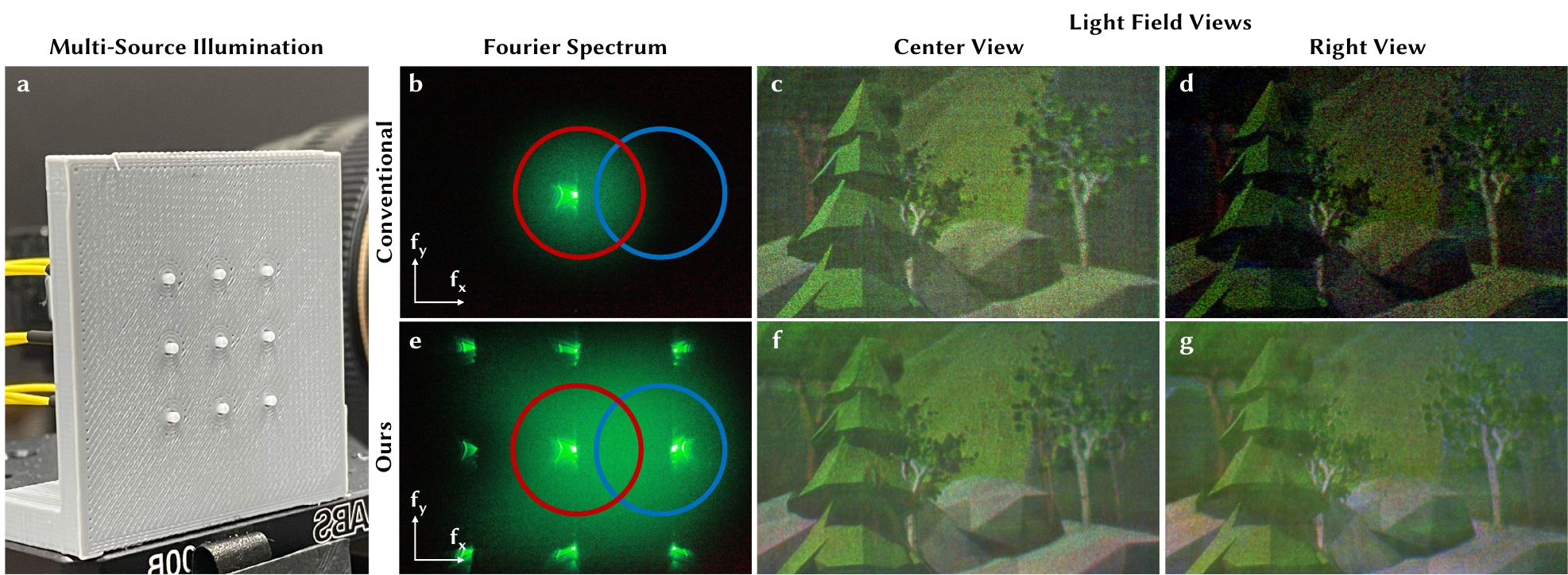}
   \caption{
Conventional holographic displays use a single laser source that provides a limited \'etendue, here visualized by a recorded spectrum that only covers a small area of the Fourier plane (b). Single-source holograms therefore only support a limited eyebox size, which means that an image can be observed only when the user's pupil (b, red) is well aligned with the eyebox (c). The image quickly degrades and fades into black as the pupil (b, blue) shifts even a small amount (d). Using multi-source illumination (a), our holographic display creates a significantly expanded coverage of addressable spatial frequencies (e) which, combined with our content-adaptive Fourier modulation strategy, achieves a large \'etendue with better image quality across an expanded eyebox (f,g). 
	}
  \label{fig:teaser}
\end{teaserfigure} 

\maketitle

\section{Introduction}
\label{sec:intro}

Holographic near-eye displays offer unique benefits to virtual and augmented reality (VR/AR) applications. For example, holographic displays can present perceptually realistic 3D images with natural parallax to the user in lightweight device form factors~\cite{maimone2017holo,Kim2022-cd, gopakumar2024ar,jang2024waveguide}. Yet, the \'etendue of holographic displays is fundamentally limited by the pixel count of the underlying spatial light modulators (SLMs), preventing current holographic near-eye displays from achieving a large field of view and eyebox simultaneously. This limitation is a fundamental barrier towards making this a practical display technology.

Increasing the pixel count of an SLM seems like the natural solution. However, developing large-area phase-only SLMs with pixel pitches matching the small feature sizes (i.e., tens of nanometers) of analog holographic films~\cite{benton2008holographic} is simply not feasible with today's hardware solutions. To overcome this problem, \'etendue expansion techniques have been described in the literature, including those based on static, high-resolution masks~\cite{kuo2020mask, park2019photon,yu2017speckle, buckley2006mask,monin2022mask,tseng2023neural}, pupil replication~\cite{kress2020waveguide}, steered or multi-source illumination~\cite{jang2018hoe,lee2020wide,jo2022binary,lee2022envelope, monin2022tilt}, and making use of higher-diffraction orders and pupil optimization \cite{shi2024ergo, schiffers2023stochastic}. However, each of these approaches has its limitations, as mask-based systems do not have sufficient degrees of freedom to achieve a high image quality, pupil replication approaches cannot create natural 3D effects and parallax over the eyebox, and steered sources are hindered by the requirement for high-speed SLMs as well as high diffraction orders (HDOs) that fundamentally limit the image quality. As a result, none of these solutions is able to achieve high-quality 3D holography with a large \'etendue.

Our work is motivated by the hypothesis that a holographic display requires sufficient degrees of freedom to achieve a large field of view and eyebox simultaneously. In the absence of an extremely high-resolution SLM, this is only achievable using steered or multi-source illumination. We thus build on the latter approach but address its major shortcomings, HDOs and symmetric illumination copies, by introducing a dynamic, programmable amplitude modulation mechanism in the Fourier plane, after the SLM. This unique optical setup allows us to extend steered / multi-source configurations such that they modulate the frequency spectrum of the display image in a content-adaptive manner. For this purpose, we leverage a stochastic optimization approach that factors a target light field into a set of time-multiplexed phase SLM and corresponding Fourier amplitude masks that are displayed in rapid succession while being integrated by the user's eye.

Using the proposed system, we demonstrate improved 3D image quality over a large \'etendue, surpassing the performances of existing approaches in both simulation and experiment. Specifically, our contributions include 
\begin{itemize}
    \item A novel optical holographic display configuration that combines a time-multiplexed phase SLM near the image plane and a dynamic amplitude SLM that controls the frequency spectrum.
    \item A computer-generated holography framework that uses stochastic optimization to factor a target light field into a set of phase--amplitude image pairs.
    \item Demonstration of improved 3D image quality among high-\'etendue holographic displays.
\end{itemize}

Our method should be clearly distinguished from Multisource Holography, a system recently proposed by \citet{grace2023multi} for speckle reduction that also leverages a multi-source laser array. In Multisource Holography, the multi-source laser and two \textit{phase-only} SLMs placed in close proximity are used to remove speckles, but the system \'etendue remains limited since the spacing between each source is relatively small. In our system, we place the laser sources much farther apart to create a high-\'etendue backlight for the phase-only SLM to greatly increase the eyebox size and place an \textit{amplitude} display at the Fourier plane.

\section{Related Work}
\label{sec:related}

\paragraph{\bf{Holographic Near-eye Displays.}} Holographic displays are a promising technology for virtual and augmented reality applications due to their unique capability to display true 3D content and significant progress has been made recently~\cite{Javidi2021-vh, Chang2020-ja, Pi2022-mr}. In particular, the advancement in computer graphics, machine learning, and computing infrastructures have enabled real-time hologram rendering based on neural networks~\cite{shi2021neural, peng2020neural}, significantly improved image quality with end-to-end optimization~\cite{peng2020neural,chakravarthula2020learned,choi2022flex}, higher light efficiency and brightness with simultaneous control of multiple wavelengths and energy-efficiency loss function~\cite{Markley2023-eg, Chao2023-wl, Kavakli2023-nd}, and thin form factors in eyeglasses-like design~\cite{maimone2017holo,jang2024waveguide, Kim2022-cd,gopakumar2024ar}. Despite offering these unique capabilities, current holographic displays fail to provide a comfortable immersive experience as they cannot simultaneously provide a wide field of view (FoV) and a sufficiently large eyebox (i.e., the region in which a user's eye perceives the displayed content). 

In a given display system, the product of the FoV and the eyebox is a constant, referred to as the {\it \'etendue}. For a holographic display, the \'etendue is directly proportional to the number of pixels in the SLM. A 1080p SLM, for instance, can either support a wide field of view (e.g., 80 degrees) with an eyebox smaller than 1~mm or vice versa. 
However, increasing the SLM resolution to the point where both large field of view and eyebox can be achieved simultaneously faces significant challenges in manufacturing, cost, and addressing speed, accuracy, and bandwidth. Instead, efforts have been made to increase the \'etendue of holographic displays without increasing SLM resolution.
The approaches fall under two categories: (i) increasing \'etendue {\it after the SLM} and (ii) increasing \'etendue {\it before} the SLM.

\paragraph{\bf{Post-SLM \'Etendue Expansion}}
The most representative method in this category is mask-based \'etendue expansion \cite{kuo2020mask, tseng2023neural, park2019photon, yu2017speckle, buckley2006mask, monin2022mask}, where a static mask at a resolution higher than the SLM is placed after the SLM to increase the diffraction angles of the SLM-modulated wavefront, thus increasing the \'etendue. However, such systems suffer from difficulties in alignment and from reduced image quality and low contrast because their effective degrees of freedom~\cite{starikov1982effective} are insufficient to synthesize a high-quality large-\'etendue wavefront. Pupil replication \cite{bigler2018expansion, bigler2019expansion, draper2019expansion, draper2021dof, draper2022expansion, jang2024waveguide, Park2018-ge,kress2020waveguide} is another popular approach. It is implemented either by putting a pupil-replicating waveguide after the SLM to replicate pupil locations at its out-coupler or using a pupil-replicating holographic optical element (HOE) as the eyepiece, effectively expanding the eyebox of the system. However, pupil-replicating displays cannot display 3D content or natural parallax across the expanded eyebox since the content within the eyebox are merely copies of the same wavefront. Higher-diffraction orders combined with pupil optimization can also be leveraged to slightly expand the eyebox in the single-source case \cite{shi2024ergo,schiffers2023stochastic}. Finally, a regular eyepiece can be replaced by a lens array to partition an unexpanded eyebox into an array of smaller chunks that cover an expanded area~\cite{Chae2023-ll, Wang2023-vf, Xia2020-sa}. However, this comes at an explicit cost of image quality and brightness nonuniformity, especially when observed with a small pupil.

\paragraph{\bf{Pre-SLM \'Etendue Expansion}}
Methods in this category modify the laser illumination to expand \'etendue either through a multi-source configuration or beam-steering. \citet{jang2018hoe} used a micro-electromechanical-system (MEMS) mirror to temporally change the laser illumination and steer the resulting pupils over a larger eye box. \citet{lee2020wide} implemented the same principle by arranging individual laser diodes into a 2D array and sequentially turning each one on to create temporal directional illumination. \citet{monin2022tilt}  implemented \textit{per-pixel} beam steering of the phase SLM by using transmissive LCD panels and polarization gratings and demonstrated that the \'etendue expansion amount scales exponentially with the number of LCD layers.
To permanently expand the \'etendue, \citet{jo2022binary} activated all illumination sources simultaneously. They introduced a random mask at the Fourier plane to break the correlation among copies in the spectrum formed by directional illuminations. This effectively eliminates duplicate images within the expanded eyebox. However, they did not demonstrate view-dependent effects across the expanded eyebox and the random mask is not content adaptive, resulting in reduced 3D realism and low image quality. Instead of using multiple laser diodes that are incoherent with each other, \citet{lee2022envelope} implemented a mutually coherent multi-laser source using a lens array. The mutually coherent sources can interfere constructively and destructively with each other, granting the hologram optimization process more degrees of freedom. However, their system requires eye tracking and a new hologram needs to be optimized for each dynamic pupil location, making the system challenging for real-time applications. 

\paragraph{\bf{Fourier Modulation.}} 
Holographic display systems often require Fourier plane filtering to remove HDOs created by the pixelated structure of the phase SLM \cite{maimone2017holo, peng2020neural, shi2021neural, Shi2022-ih}. However, it is not straightforward to apply this to a multi-source or beam steering setting since the directional illuminations create shifted copies of the wavefront from normally incident illumination and the associated HDOs in the Fourier domain. When using beam-steering, the filter position needs to be dynamically adjusted to block the HDOs of the shifted wavefront. To achieve this, \citet{lee2020wide} placed a programmable polarization shutter at the Fourier plane and synchronized the laser sources to filter out the HDOs. However, when using multi-source illumination for \'etendue expansion, HDOs associated with one illumination intermingle with the shifted wavefront of another illumination at the Fourier domain, making it impossible to separate. It is therefore crucial to model HDOs to precisely characterize how they contribute to different angular views. Explicit modeling of HDOs has been demonstrated for 2D images~\cite{gopakumar2021unfiltered} and 3D focal stacks~\cite{Kim2022-cd, shi2024ergo}, but not for 4D light fields under multi-source illumination.

\bigskip

Inspired by \citet{jo2022binary}, we employ a multisource laser illumination and additionally place a dynamic amplitude SLM at the Fourier plane to enable content adaptive modulation and time multiplexing. We jointly optimize the patterns for the phase and amplitude SLMs to reproduce a 4D light field rendered over an expanded eyebox. We also explicitly model the HDOs and demonstrate notable improvement in image quality and contrast. Collectively, this new hardware and software co-design enables a dynamic view-dependent holographic display with a large eyebox.

\section{Method}
\label{sec:method}
\newcommand{\usource}{u_\text{src}}
\newcommand{\uslm}{u_\text{slm}}
\newcommand{\uz}{u_\text{z}}
\newcommand{\utarget}{u_\text{target}}
\newcommand{\upupil}[1]{u_{\text{pupil#1}, p}}
\newcommand{\Fuslm}{\mathcal{U}_\text{slm}}
\newcommand{\Futarget}{\mathcal{U}_\text{target}}
\newcommand{\Fupupil}[1]{\mathcal{U}_{\text{pupil#1}, p}}
\newcommand{\filter}{\mathcal{P}}
\newcommand{\pupil}{\mathcal{M}}
\newcommand{\fourier}{\mathcal{F}}
\newcommand{\transfer}{\mathcal{H}}
\newcommand{\invfourier}{\mathcal{F}^{-1}}
\newcommand{\fx}{f_x}
\newcommand{\fy}{f_y}
\newcommand{\cx}[1]{c_{x, #1}}
\newcommand{\cy}[1]{c_{y, #1}}
\newcommand{\numpixelxslm}{N_{x, slm}}
\newcommand{\numpixelyslm}{N_{y, slm}}
\newcommand{\numpixelxdmd}{N_{x, dmd}}
\newcommand{\numpixelydmd}{N_{y, dmd}}

In this section, we first review the conventional single-source holographic image formation model before introducing the multi-source image formation model of our system.

\begin{figure}[t]
    \centering
    \includegraphics[width=1.0\linewidth]{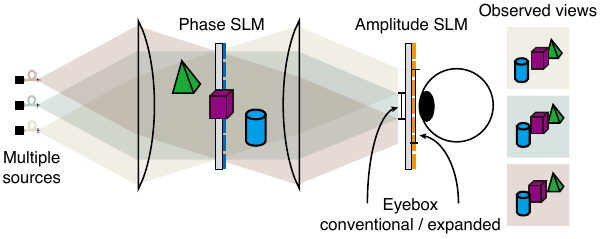}
    \caption{\textbf{System Architecture.} Multiple mutually incoherent sources illuminate a fast phase-only SLM, creating a high-\'etendue backlight. An additional amplitude display is placed at the Fourier plane to remove ghost image artifacts created by HDOs and the multisource illumination. The phase and amplitude patterns are optimized jointly for a target light field in a content adaptive manner. We illustrate our design following the style of the schematic in \cite{grace2023multi} for easier comparison.}
\label{fig:schematic_conceptual}
\end{figure}

\subsection{Single-Source Holographic Image Formation Model} 

For on-axis Fresnel holography, a collimated beam from a laser source illuminates an SLM with a normally incident, coherent field $\usource$. The SLM imparts a spatially varying phase delay $\phi$ to the field which propagates a distance $z$ along the optical axis. The wavefront at this plane can be mathematically described using the angular spectrum method (ASM)~\cite{goodman2005fourier} as a function of the phase pattern and distance from the SLM:

\begin{equation}
\begin{split}
    & f \left(\phi, z \right) = \invfourier \left\{ \fourier \left\{ e^{i\phi \left(x, y \right)} \usource \left(x, y \right) \right\} \cdot \transfer \left(\fx, \fy; z \right) \right\} \\
    &\transfer \left(\fx, \fy; z \right) = \begin{cases}
    e^{i\frac{2\pi}{\lambda} z \sqrt{1- \left( \lambda \fx \right) ^2- \left( \lambda \fy \right) ^2} } & \text{if} \, \sqrt{\fx^2+\fy^2} < \frac{1}{\lambda}.\\
    0 & \text{otherwise}.
\end{cases}
\end{split}
\label{eq:asm}
\end{equation}
Here, $\lambda$ is the wavelength of light, $x, y$ are the spatial coordinates on the SLM, $\fx, \fy$ are the frequency coordinates, 
and $\transfer$ is the transfer function of the ASM. The operator $f$ models free-space propagation between the parallel SLM and target planes separated by a distance $z$. For notational convenience, we omit the dependence
of the fields on $x$ and $y$. The intensity generated by a holographic display at a distance $z$ in front of the SLM is therefore $\left| f \left(\phi, z \right) \right|^2$. If a high-speed SLM is available, a time-multiplexed variant of the image formation is $\sum_{t=1}^T \left| f \left(\phi^{(t)}, z \right) \right|^2 \!\! / T$, where $T$ phase SLM patterns $\phi^{(t)}, t = 1, \ldots, T$ are rapidly displayed in sequence, and the resulting intensities are averaged by the users' eye~\cite{choi2022flex}.

\subsection{Multi-Source Holographic Image Formation Model with Fourier Modulation}
To extend the single-source image formation model to our system, we modify the formulation to incorporate off-axis collimated illumination traveling in direction
$\mathbf{k} = (k_x, k_y, k_z)$ and a programmable amplitude mask $\filter$ at the Fourier plane of the holographic display system. This results in the model

\begin{align}
    & f^{(j)} \left(\phi, \filter, z \right) = 
    \invfourier \left\{ \Fuslm^{(j)} \left(\fx, \fy; \phi \right) \cdot \transfer \left(\fx, \fy; z \right) \cdot \filter(\fx, \fy) \right\} \nonumber \\     
    & \Fuslm^{(j)} \left(\fx, \fy ; \phi \right) = \fourier \left\{ e^{i\phi \left(x, y \right)} \usource^{(j)} \left(x, y \right) e^{i \, \mathbf{k}^{(j)} \cdot \mathbf{x}}\right\}
\label{eq:multi}
\end{align}
where $j$ is the index of the source, 
and $\usource^{(j)} \left(x, y \right)$ is the complex-valued field modeling any deviations in amplitude and phase of source $j=1, \dots, J$ from a perfect plane wave $e^{i \, \mathbf{k}^{(j)} \cdot \mathbf{x}}, \mathbf{x} = (x,y,z)$. Moreover, $\usource^{(j)} \left(x, y \right)$ can optionally also include per-source, time-dependent modulation, such as switching individual lasers on and off. In our setup, we do not consider this case and assume that all sources are turned on at all times. Please refer to the supplemental material for more discussions about the generalized configuration with amplitude-controllable laser sources $\usource^{(j)} \left(x, y \right)$.

\subsection{Stochastic Optimization of Light Field Holograms}

To reconstruct a light field, we use gradient descent to optimize a set of time-multiplexed phase patterns $\phi^{(t)}$ and corresponding Fourier masks $\filter^{(t)}$ by minimizing the following objective:
\begin{equation}
    \underset{\left\{\phi^{(t)}, \filter^{(t)} \right\}}{\textrm{minimize}} \left  \lVert s \sqrt{\frac{1}{T}\sum_{t=1}^{T} \sum_{j=1}^{J} |\text{H2LF}\left( f^{(j)} \left(\phi^{(t)}, \filter^{(t)}, z\right) \right)}|^2 - \textrm{lf}_\text{target} \right \rVert 
\label{eq:h2lf}
\end{equation}
where $s$ is a scale factor, $\textrm{lf}_\text{target}$ is the amplitude of the target light field, and $\text{H2LF}$ is a hologram-to-light field transformation, such as the Short-Time Fourier Transform (STFT)~\cite{zhang2009wigner,padmanaban2019olas}. 

The memory consumption of the above optimization problem is huge due to time multiplexing, multiple sources, and the explicit modeling of HDOs. Therefore, it is impractical to realize $\text{H2LF}$ using the STFT since it reconstructs a whole light field and the memory consumption would explode for a dense $\text{lf}_\text{target}$. To solve this problem, we devise a $\textit{stochastic}$ version of Eq. \ref{eq:h2lf} that allows us to optimize a single light-field view rather than a full light field in each iteration of the optimization routine. 

For this purpose, we randomly chose a view $p$ of the target light $\text{lf}^{(p)}_\text{target}$ in each iteration and run a gradient descent step of Eq.~\ref{eq:h2lf}. A binary pupil mask $\pupil^{(p)}$ in the Fourier plane in the hologram-to-light field transform is applied to reconstruct one specific view as 
\begin{equation}
\begin{split}
    & \text{H2LF}^{(p)}\left( f^{(j)} \left(\phi^{(t)}, \filter^{(t)}, z\right) \right) = f^{(j)} \left(\phi^{(t)}, \filter^{(t)} \cdot \pupil^{(p)}, z\right), \\
    & \pupil^{(p)}(\fx, \fy) = \begin{cases}
        1, & \text{if} \, (\fx - \cx{p})^2 + (\fy - \cy{p})^2 \leq r_p^2, \\
        0, & \text{otherwise}
    \end{cases}
\end{split}
\label{eq:pupil}
\end{equation}
where $\pupil^{(p)}$ is a binary pupil mask in the Fourier plane, $r_p$ is the radius of the pupil and $\cx{p}, \cy{p}$ are the spatial coordinates of the center of the pupil. This procedure is similar to the pupil-supervision techniques described in \cite{shi2024ergo, schiffers2023stochastic, chakravarthula2022pupil}. Please refer to the supplemental material for more details on our stochastic light field optimization procedure.

\subsection{Implementation Details}

Since we are using a highly-quantized 4-bit phase SLM, the quantization of pixel values needs to be taken into consideration. Such quantization constraints can be enforced using techniques described in prior work~\cite{choi2022flex}. Higher diffraction orders (HDOs) are modeled using the wave propagation model described in \cite{gopakumar2021unfiltered}. We use PyTorch to implement all our algorithms and run optimization.

In all our experiments, the radius $r_p$ of the pupils are set to be 2 mm, resulting in a 4 mm diameter pupil. 81 pupils are equally spaced in the Fourier plane (eyebox plane), where each pupil corresponds to a single view in a $9 \times 9$ light field. The illumination directions of the multisource laser are set such that they match the diffraction angle of the $\pm 1^{\text{st}}$ higher diffraction orders of the blue wavelength. This allows the blue spectrum copies to be perfectly tiled in the Fourier plane, while removing the gaps between the red and green wavelength spectrum copies. Please see the supplemental materials for detailed discussion on the choosing the appropriate illumination angles.

\section{Analysis}
\label{sec:analysis}

\newcommand{\fl}{g}

\subsection{Optical System Analysis}
The \textit{\'etendue} $G$ of a display is defined as the product of the display area with the solid angle of emitted light:
\begin{equation}
    G = 4A\text{sin}^2\theta,
\end{equation}
where $A$ is the display area and $2\theta$ is the solid angle of the emission cone of each display pixel. \'Etendue is conserved through reflections, refractions, and free space propagation in an optical system. When illuminated with a normal incidence light of wavelength $\lambda$, the diffraction angle $\theta_\text{SLM}$ of an SLM with pixel pitch $p$ can be expressed as $\theta_\text{SLM} = \pm \text{sin}^{-1}\frac{\lambda}{2p}$.
   
For an SLM of physical size $L_x \times L_y$, its \'etendue $G_\text{SLM}$ can be expressed as:
\begin{equation}
    \begin{gathered}
    G_\text{SLM} = 4L_x L_y\text{sin}^2\theta_\text{SLM} = \lambda^2 N_x N_y    
    \end{gathered}
\end{equation}
where $N_x \times N_y$ is the pixel resolution of the SLM. This means that the \'etendue of a holographic display is directly proportional to the number of pixels of the SLM.

\begin{figure}[t]
    \centering
    \includegraphics[width=\linewidth]{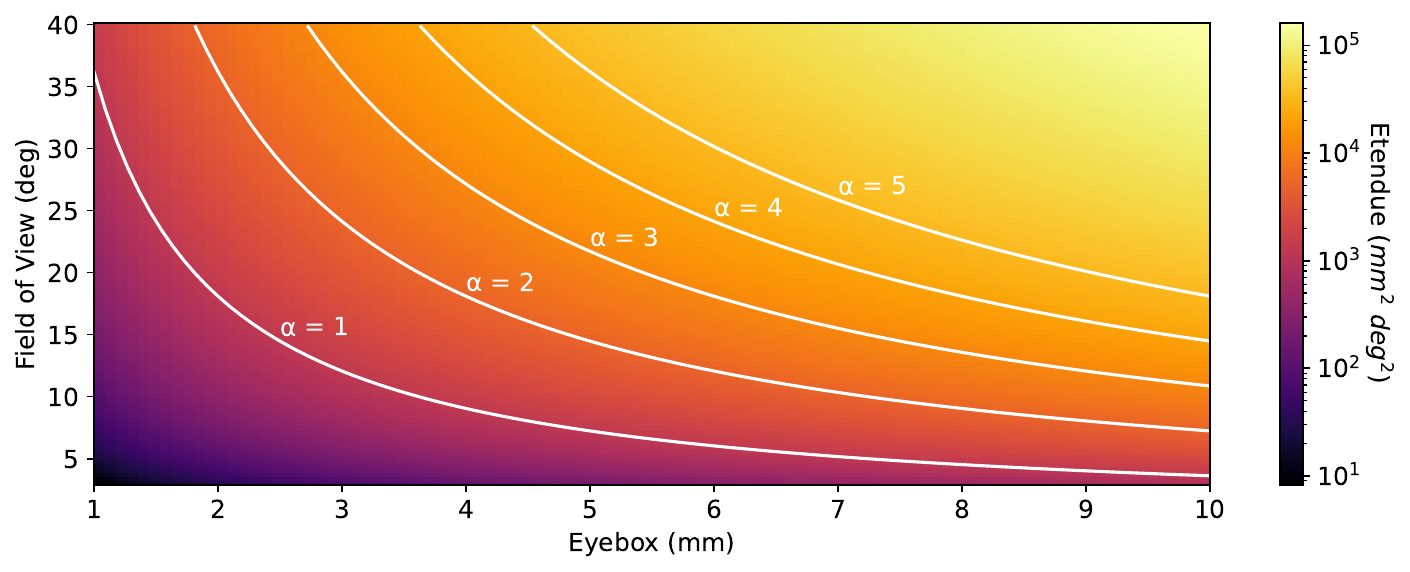}
    \caption{\textbf{Tradeoff between 2D field of view (FoV) and eyebox size}. Each white line represents the fixed \'etendue of a holographic display system illuminated by a grid of $\alpha \times \alpha$ sources with different $\alpha$ values. We show the \'etendue of the systems in log-scaled color maps. As the number of sources increase, the \'etendue of the system also increases.}
    \label{fig:fov_eyebox_tradeoff}
\end{figure}

\begin{figure*}[t]
    \centering
    \includegraphics[width=\linewidth]{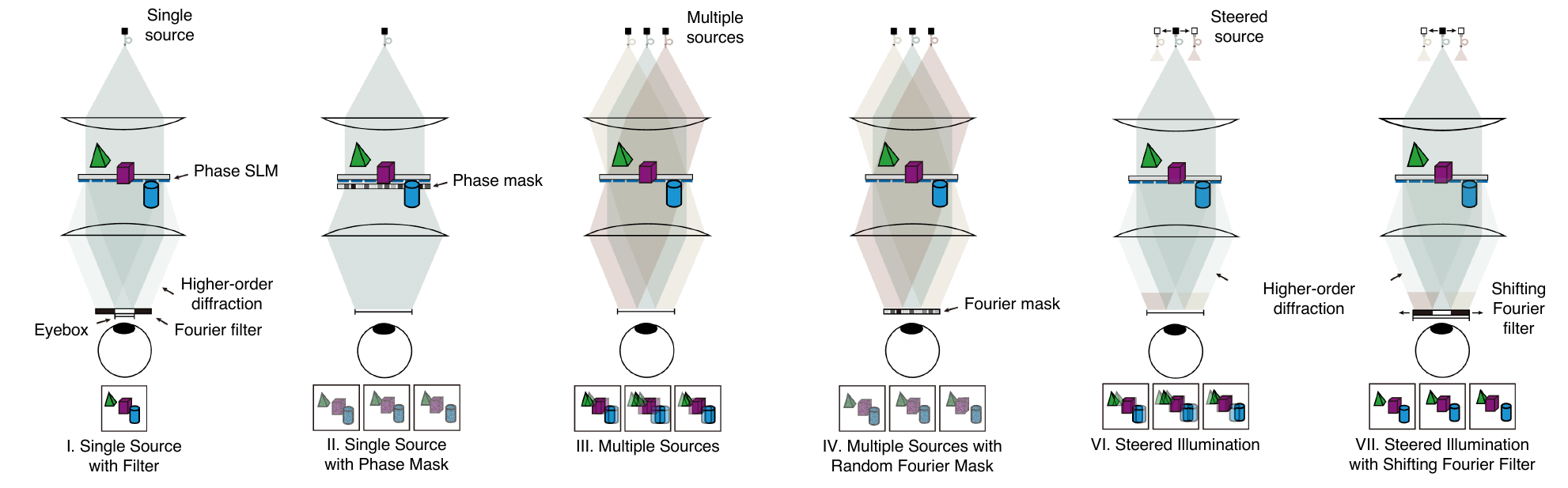}
    \caption{\textbf{Illustration of baseline display configurations.} Single-source configurations (I, II) trade image quality off for \'etendue expansion factor (II). Multi-source approaches that use all sources simultaneously (III, IV) benefit from a high-\'etendue ``backlight'' but operate within limited effective degrees of freedom, which also makes it challenging to achieve a high image quality. Steered illumination approaches (VI, VII) sequentially illuminate the system from different directions and require high-speed SLMs. Without a Fourier mask, the image quality achieved by these systems is also limited by high diffraction orders (VI). We illustrate our design following the style of the schematic in \cite{grace2023multi} for easier comparison.}
    \label{fig:baseline}
\end{figure*}

In a Fresnel holography display system, the 1D field-of-view (FoV) and eyebox size $w$ can be expressed as follows:
\begin{equation}    
        \text{FoV} = 2\text{tan}^{-1} \left(\frac{L}{2 \fl} \right) = 2\text{tan}^{-1} \left(\frac{Np}{2 \fl} \right), \quad \,\,
        \text{w} = \frac{\fl \lambda}{p},
\end{equation}
where $L, N$ are the size of the SLM and the number of SLM pixels in the $x$ or $y$ axis, respectively, and $\fl$ is the eyepiece focal length. Under paraxial assumptions ($\theta \approx \text{sin}\theta \approx \text{tan}\theta)$, we see that the product of the 2D FoV and eyebox of a holographic display system is exactly the \'etendue of the system:
\begin{equation}
\begin{gathered}
    \text{FoV}_x \cdot \text{FoV}_y \cdot w^2 = \\
    2\text{tan}^{-1} \left(\frac{N_xp}{2 \fl} \right) \cdot 2\text{tan}^{-1} \left(\frac{N_yp}{2 \fl} \right) \cdot \frac{\fl^2\lambda^2}{p^2} \approx \lambda^2 N_x N_y = G_\text{SLM}
\end{gathered}
\end{equation}
This implies that there is an inherent tradeoff between the FoV and the eyebox of a holographic display.

When the SLM is illuminated with a grid of $\alpha \times \alpha$ off-axis, directional illuminations, the system eyebox is expanded due to shifted copies of the original spectrum. Specifically, if the directional illumination is selected such that the illumination direction matches the higher-order diffraction angles, the system 1D eyebox is exactly expanded by $\alpha$ while the FoV remains the same, resulting in an expanded 1D eyebox size of $\text{w} = \frac{\alpha f\lambda}{p}$. Therefore, the 2D \'etendue of the system is expanded by a factor of $\alpha^2$. 

We show how the FoV and eyebox size relates to the required number of sources in Fig. \ref{fig:fov_eyebox_tradeoff}. We assume an SLM pixel pitch of $10.8 \; \mu$m and resolution of $1000 \times 1000$ and laser wavelength of $632.8$ nm. The FoV and eyebox size move along each white line in opposite directions as we vary the eyepiece focal length $\fl$ while the system \'etendue remains fixed. As we increase the number of sources, the \'etendue of the system also increases, as the white lines move further towards to upper-right of the plot.

\subsection{Baseline Configurations}

We next discuss a number of holographic display system configurations that serve as baselines to our proposed design shown in Fig. \ref{fig:schematic_conceptual}. Illustrations of these baselines are shown in Fig.~\ref{fig:baseline}.

\paragraph{I. Single Source with Fourier Filter.} The conventional holographic display setup with a single laser source and a Fourier filter to block HDOs, including \cite{peng2020neural, shi2021neural, choi2022flex, maimone2017holo}. Such systems suffer from small \'etendue and non-uniform brightness across the eyebox.

\paragraph{II. Single Source with Phase Mask.} A high-res\-o\-lu\-tion phase mask is placed in front of the SLM to increase the diffraction angle of the SLM and therefore increase the \'etendue of the system. The phase masks can be random \cite{kuo2020mask, buckley2006mask, park2019photon, yu2017speckle} or optimized \cite{tseng2023neural,monin2022mask}. These approaches have been shown to expand the \'etendue at the cost of decreased image quality and contrast.

\paragraph{III. Multiple Sources.} Multiple mutually incoherent lasers illuminate the SLM from different angles \emph{simultaneously}. Due to the absence of a Fourier filter, the frequency spectrum contains multiple shifted, potentially overlapping copies of the same hologram. These constraints limit this system's capability to perfectly reconstruct a light field.


\paragraph{IV. Multiple Sources with Fixed Random Fourier Mask} Multiple lasers illuminate the SLM from different angles \emph{simultaneously} while a fixed random mask is placed at the Fourier plane to break the correlation between the image copies, as demonstrated by Jo et al.~\shortcite{jo2022binary}. Time multiplexing and content-adaptive filtering are not feasible since the random masks are custom-printed and fixed.

\paragraph{V. Multiple Sources with Dynamic Fourier Filter (ours).} Multiple lasers illuminate the SLM from different angles \emph{simultaneously} while an amplitude SLM is placed at the Fourier plane. The amplitude SLM can be dynamically refreshed and is synchronized with the phase SLM, allowing for time-multiplexed and content-adaptive Fourier modulation. We additionally compare with a generalized configuration $\text{V}^*$ where the amplitude of the laser sources are controllable rather than fixed. More discussions on this generalized configuration can be found in the supplement.

\paragraph{VI. Steered Illumination.} Multiple individually controllable sources or a single, swept source illuminate the SLM from different angles without Fourier filtering, including \cite{chang2019maxweillian, an2020slim, jang2018hoe, reichelt2010holo, xia2023steer}. Reduced image quality due to HDOs remains an issue due to the lack of filtering. Furthermore, such methods only apply to very high-speed SLMs, because each source is sequentially turned on or steered in sequence.

\paragraph{VII. Steered Illumination with Shifting Fourier Filter.} Multiple individually controllable sources or a single, swept source illuminate the SLM from different angles while a synchronizable, dynamic filter is placed at the Fourier plane to filter out the HDOs. One example is the steered illumination system described in \cite{lee2020wide}. This method is still sequential in nature, as each source is turned on one at a time, and requires a very high-speed SLM.

\subsection{Assessment}

\begin{figure*}[p]
    \centering
    \includegraphics[width=0.98\textwidth]{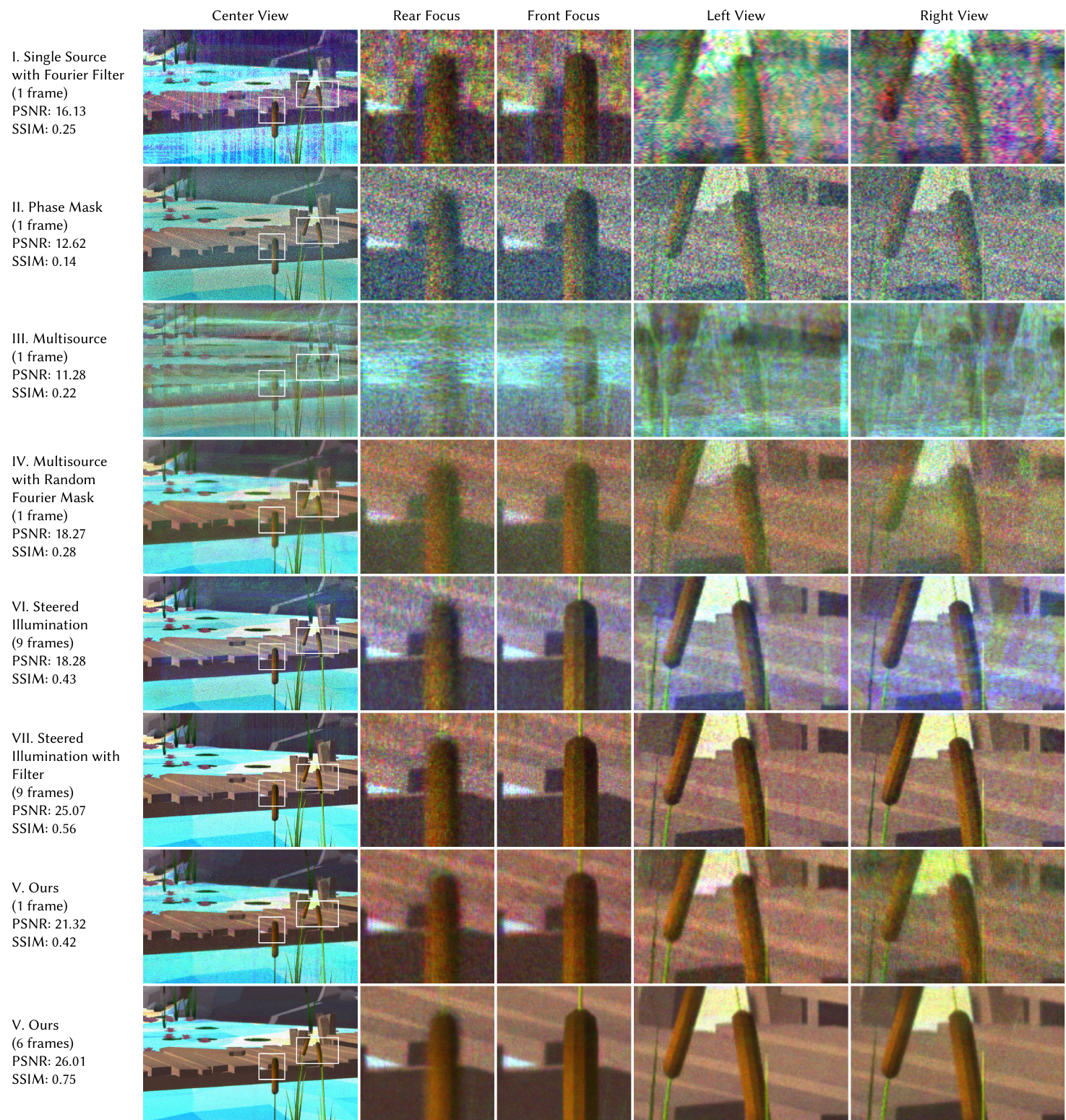}
    \caption{\textbf{Comparison of hardware configurations using simulated reconstruction.} Here, we compare different \'etendue-expanded holographic display configurations, including the conventional setup with light-field supervision (1st row), phase masks~\cite{kuo2020mask} (2nd row), multisource (3rd row), multisource with random Fourier mask~\cite{jo2022binary} (4th row), steering (5th row), steering with filter (6th row), and ours with 1-frame and 6-frame time multiplexing (multisource with content-adaptive dynamic Fourier modulation, 7--8th rows). For each configuration, we present the central view (1st column) and insets at two focal slices (rear and front) and two different viewpoints (left and right) in the next four columns. Quantitative evaluations are included as PSNR (dB)/SSIM on the left. Note that all methods use the same, large-baseline target light field for supervision, which degrades the quality of the single-source configuration (I) because it simply does not support such a large eyebox.}
    \label{fig:simulation}
\end{figure*}

Table \ref{tab:baseline_comparison_recon} and Figure~\ref{fig:simulation} shows the light field reconstruction performance of different baseline configurations in simulation. For configurations with multiple sources (III, IV, V, V*, VI, and VII), we consider a $3\times3$ grid of sources. We simulate a $800 \times 1280$ phase SLM for single-SLM configurations (I, II, III, VI, VII) and an additional $20 \times 20$ Fourier display for configurations IV, V, and $\text{V}^*$. We run our optimization algorithm on all configurations to reconstruct a $9 \times 9$ light field. Additionally, we optimize the single source configuration to reconstruct a smaller, $3 \times 3$ light field in Table \ref{tab:baseline_comparison_recon}. Time multiplexing is not used for configurations I, II, III, and IV. Please refer to the supplemental material for additional discussions on the optimization parameters, degrees of freedom of the system, and ablation studies on the resolution of the Fourier display.

The naive single-source configuration (I) is able to reconstruct a small $3 \times 3$ light field, but fails to reconstruct a larger-baseline $9 \times 9$ light field and cannot support uniform brightness across the expanded eyebox (Fig.~\ref{fig:teaser}, b--d). Mask-based \'etendue expansion techniques (II) reconstruct low-contrast and speckly images. By using multiple sources (III), the eyebox is expanded but the light field reconstruction quality is poor due to copies created by multiple sources and HDOs. Introducing a fixed random mask at the Fourier plane (IV) improves image quality, although the improvement is limited due to the lack of time multiplexing and content-adaptive Fourier mask optimization. Steered illumination options (VI, VII) achieve decent image quality, however both configurations reconstruct speckly images due to HDOs and can only be implemented using high-speed SLMs due to the large number of required time-multiplexed frames.

Our methods (V, $\text{V}^*$) achieves the best image reconstruction quality when using one frame and is better than the steered illumination baseline (VII) while using fewer frames (6 vs. 9). This is achieved through time multiplexing and our novel content-adaptive Fourier modulation optimization framework. Our method successfully removes the copies created by multiple sources and HDOs, reconstructing clean and speckless light field views. More importantly, a minimal increase in degrees of freedom in the Fourier plane (a low-resolution $20 \times 20$ Fourier display) is sufficient to achieve good image quality, and we perform extensive experiments to validate this claim in the supplemental material. Finally, although our generalized configuration $\text{V}^*$ with amplitude-controllable sources achieves the best quantitative image quality, the improvement is marginal (<0.5 dB in terms of PSNR) and suffers from a much higher system complexity. Hence, we opted for configuration V for our hardware implementation.

\begin{table}[t]
   \footnotesize
  \centering
  \renewcommand{\arraystretch}{2}
  \begin{tabular}{@{}P{0.60 \linewidth}P{0.35 \linewidth}@{}}
    \toprule
    \makecell[l]{Configuration} & \makecell[l]{Image Quality (PSNR/SSIM)} \\
    \midrule
    \makecell[l]{I. Single Source with Fourier Filter \\ 
    \cite{peng2020neural, choi2022flex, shi2021neural}} & \makecell[l]{46.23 / 0.97 (2D image) \\ 22.43 / 0.46 ($3 \times 3$ light field) \\ 16.60/ 0.23 ($9 \times 9$ light field)} \\
    \makecell[l]{II. Single Source with  Phase Mask \cite{kuo2020mask}} & \makecell[l]{14.61 / 0.15} \\
    \makecell[l]{III. Multiple Sources} & \makecell[l]{13.35 / 0.24} \\
    \makecell[l]{IV. Multiple Sources with Fixed Random Fourier \\ Mask \cite{jo2022binary}} & \makecell[l]{18.63 / 0.26} \\
    \makecell[l]{V. Multiple Sources with Dynamic Fourier Filter \\(ours)} &  \makecell[l]{21.05 / 0.39 (1 frame) \\ 24.65 / 0.66 (6 frames) \\ 25.17 / 0.71 (9 frames)} \\
    \makecell[l]{$\text{V}^*$. Multiple Sources with Dynamic Fourier Filter \\ with Laser Amplitude Control} &  \makecell[l]{ 21.43 /  0.42 (1 frame) \\ 25.02 / 0.69 (6 frames) \\ 25.66 / 0.74 (9 frames)} \\
    \makecell[l]{VI. Steered Illumination} & \makecell[l]{20.02 / 0.43 (9 frames)}  \\
    \makecell[l]{VII. Steered Illumination with Shifting Fourier Filter \\ \cite{lee2020wide}} & \makecell[l]{23.95 / 0.51 (9 frames)} \\
    \bottomrule
  \end{tabular}
  \caption{\textbf{Baseline comparisons in simulation.} Light field reconstruction performance of different baseline configurations in terms of PSNR/SSIM. We run our optimization algorithm on all configurations and six test scenes to reconstruct $9 \times 9$ light fields and report the average reconstruction performance. We run additional optimizations on the single source configuration to reconstruct a 2D image and a smaller $3\times 3$ light field. Our method achieves the best reconstruction quality for the full light field while using fewer frames. Per-scene PSNR/SSIM can be found in the supplemental materials.}
  \label{tab:baseline_comparison_recon}
\end{table}

\section{Experimental Results}
\label{sec:experiments}

\begin{figure}[t!]
    \centering
    \includegraphics[width=\columnwidth]{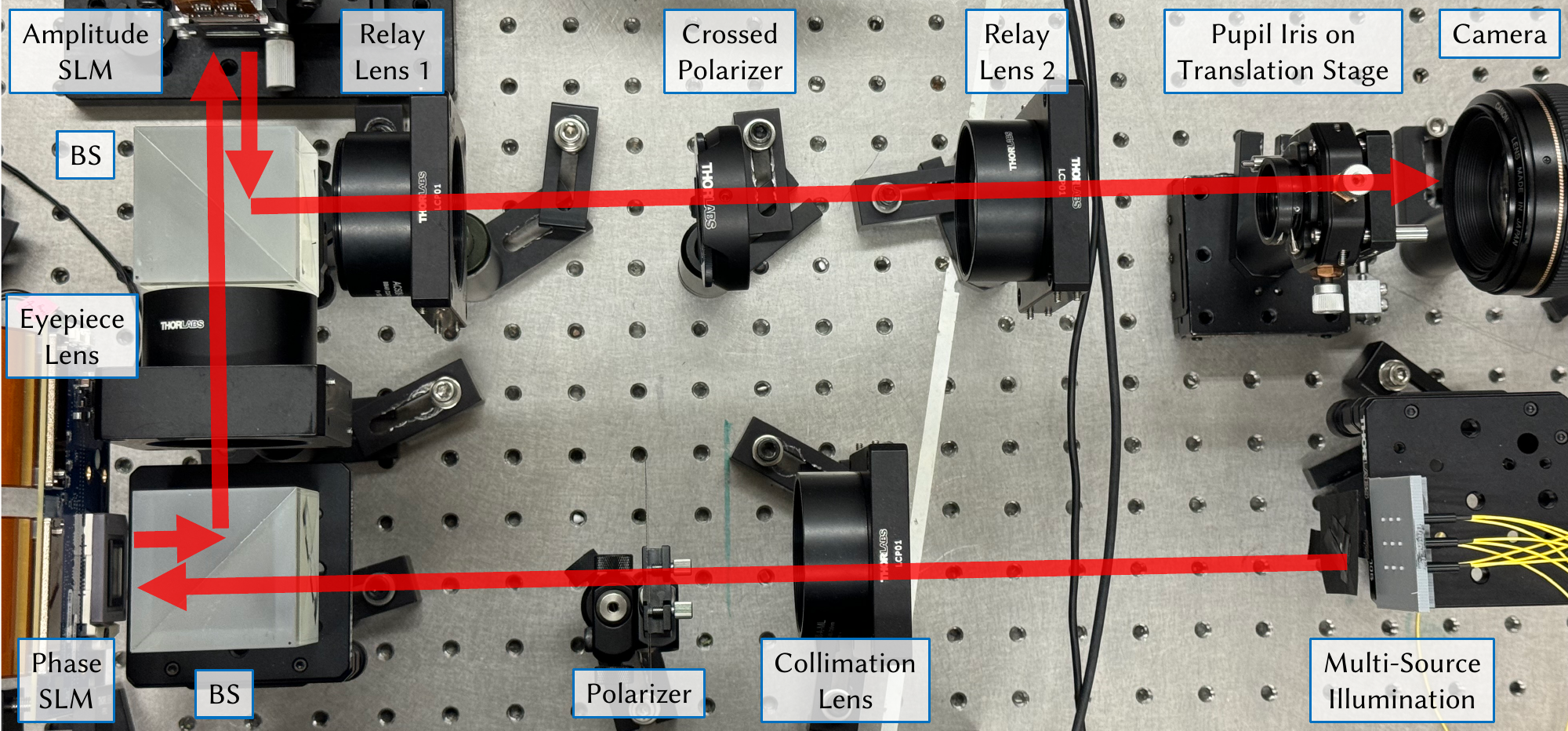}
    \caption{\textbf{Photograph of our multi-source holographic display prototype.} The propagation path is illustrated in red and components are labeled.}
  \label{fig:hw_setup}
\end{figure}

\paragraph{\bf{Hardware Implementation}} We implement the proposed 3D holographic display design and evaluate our algorithms on the system. The hardware setup and the optical path are shown in Fig.~\ref{fig:hw_setup}. We implement our multi-source laser by cascading multiple 1:4 fiber splitters (Thorlabs TWQ560HA) and arranging 9 customized fiber tip outputs into a $3\times3$ array, which is then held together using a custom-printed 3D mount. The spacing between each source is 8.17 mm and a 200 mm lens is used to collimate the multi-source laser. Each collimated source field is, therefore, incident on the phase SLM with a $2.34^{\circ}$ incident angle. We use a TI DLP6750Q1EVM phase SLM for phase modulation and a 1080p SiliconMicroDisplay liquid crystal on silicon (LCoS) display for Fourier amplitude modulation. A 75 mm Fourier transform lens is used to image the spectrum of the phase-modulated wavefront onto the amplitude SLM. Our final design has a diagonal field-of-view (FoV) of 7.78 degrees and an eyebox size of 8.53 mm $\times$ 8.53 mm. A FLIR Grasshopper 2.3 MP color USB3 vision sensor paired with a Canon EF 50mm f/1.4 USM camera lens is used to capture all experimental results. Please refer to the supplemental material for additional details on the degrees of freedom of our system and the relevant optimization parameters.

\paragraph{\bf{Experimental Capture Details}} 
To capture light field views, we place pupil masks at the Fourier plane to mimic the movement of the user's eyes, which is a technique used in prior works \cite{schiffers2023stochastic,shi2024ergo}. We implement this with a Thorlabs SM1D12 adjustable iris on a translation stage at the Fourier plane. To capture focal stacks, we center the pupil at the Fourier plane and adjust the camera focus to capture images at different depths.

\paragraph{\bf{Assessment}} 
Experimentally captured results are shown in Figs.~\ref{fig:teaser}, \ref{fig:experimental}, Table \ref{tab:baseline_comparison_experiment}, and in the supplemental material. The PSNR and SSIM values are averaged across all captured light field views. We observe the same trends as predicted by our simulations both quantitatively and qualitatively: the single-source configuration only supports a limited eyebox and suffers from severe brightness falloff at peripheral viewpoints; 3D multi-source holography without a Fourier filter cannot achieve a high image quality due to the copies created the multiple sources; a static random mask placed in the Fourier plane only provides limited degrees of freedom and suffers from low contrast; our approach without time multiplexing (i.e., 1 frame) improves the quality over the random mask as it optimizes the amplitude mask pattern in a content-adaptive manner; our method with 6-frame time multiplexing achieves the highest image quality with the largest amount of empirically observed parallax.

\begin{table}[t]
   \footnotesize
  \centering
  \renewcommand{\arraystretch}{2}
  \begin{tabular}{@{}P{0.60 \linewidth}P{0.35 \linewidth}@{}}
    \toprule
    \makecell[l]{Configuration} & \makecell[l]{Image Quality (PSNR/SSIM)} \\
    \midrule
    \makecell[l]{III. Multiple Sources} & \makecell[l]{12.72 / 0.19} \\
    \makecell[l]{IV. Multiple Sources with Fixed Random Fourier \\ Mask \cite{jo2022binary}} & \makecell[l]{12.46 / 0.17} \\
    \makecell[l]{V. Multiple Sources with Dynamic Fourier Filter \\(ours)} &  \makecell[l]{13.83 / 0.23 (1 frame) \\ \textbf{14.43} / \textbf{0.40} (6 frames) } \\ 
    \bottomrule
  \end{tabular}
  \caption{\textbf{Experimentally captured baseline comparisons.} Experimentally captured light field reconstruction performance of different baseline configurations in terms of PSNR/SSIM, averaged across all six test scenes. Our method achieves the best experimental image quality. Per-scene PSNR/SSIM can be found in the supplemental materials.}
  \label{tab:baseline_comparison_experiment}
\end{table}

\section{Discussion}
\label{sec:discussion}
In summary, we present a novel hardware system for \'etendue expansion and an algorithmic framework for 4D light-field-supervised computer-generated holography. The hardware system includes a multi-source laser array to create a large-\'etendue coherent backlight for the phase SLM and an amplitude SLM for dynamic Fourier-amplitude modulation. The algorithmic framework includes the joint optimization of time-multiplexed amplitude SLM and phase SLM patterns and a memory-efficient, stochastic light field supervision procedure to create 4D light field holograms. We compare our method with a number of \'etendue expansion baselines and verify in simulation and experimentally that our system achieves the highest-quality light field reconstruction results for large \'etendue settings. 

\paragraph{\bf{Limitations and Future Work}} 
We demonstrate our results on a benchtop display setup but futher efforts are required to miniaturize this system. Currently, our multisource laser array is implemented using bulky fiber splitters and could be miniaturized using nanophotonic phased arrays~\cite{sun2013large}. Folding the propagation distance of holograms using optical waveguides could further remove the need of beam splitters and subsequently shrink the form factor, as demonstrated in \cite{jang2024waveguide, yeom2021voxel, lin2018wedge, lin2020waveguide}. We illustrate potential compact designs in the supplemental material. The frame rate of our system is limited by our amplitude display (240 Hz native frame rate). This translates to a $\sim$13.33 Hz frame rate when operating in color-sequential mode with 6-frames time-multiplexing. The frame rate can be improved by using more advanced LCoS displays with frame rate > 720 Hz \cite{lazarev2017lcos}. Real-time synthesis of light field holograms are necessary for practical holographic displays, but is not currently not supported by our system. Extending recent neural network-based hologram synthesis methods \cite{shi2021neural, Shi2022-ih, yang2022diffeng} to work for 4D light field holograms would be an interesting future direction. Finally, we did not attempt to calibrate a neural network--parameterized wave propagation model of our prototype display system, which has been demonstrated to significantly improve experimentally captured holographic image quality for other types of optical configurations~\cite{peng2020neural,choi2022flex}.

\paragraph{\bf{Conclusion}} The novel hardware design and algorithmic framework presented in this work improves the \'etendue of holographic displays and allows for light field holograms synthesis with improved image quality. These help make holographic displays a more practical technology for augmented and virtual reality applications. 

\begin{acks}
We thank Grace Kuo for helpful advice regarding the implementation of the multisource laser setup. Brian Chao is supported by the Stanford Graduate Fellowship and the NSF GRFP. Manu Gopakumar is supported by the Stanford Gradudate Fellowship. Suyeon Choi is supported by the Meta Research PhD Fellowship.
\end{acks}

\clearpage

\begin{figure*}[p]
    \centering
    \includegraphics[width=0.96\textwidth]{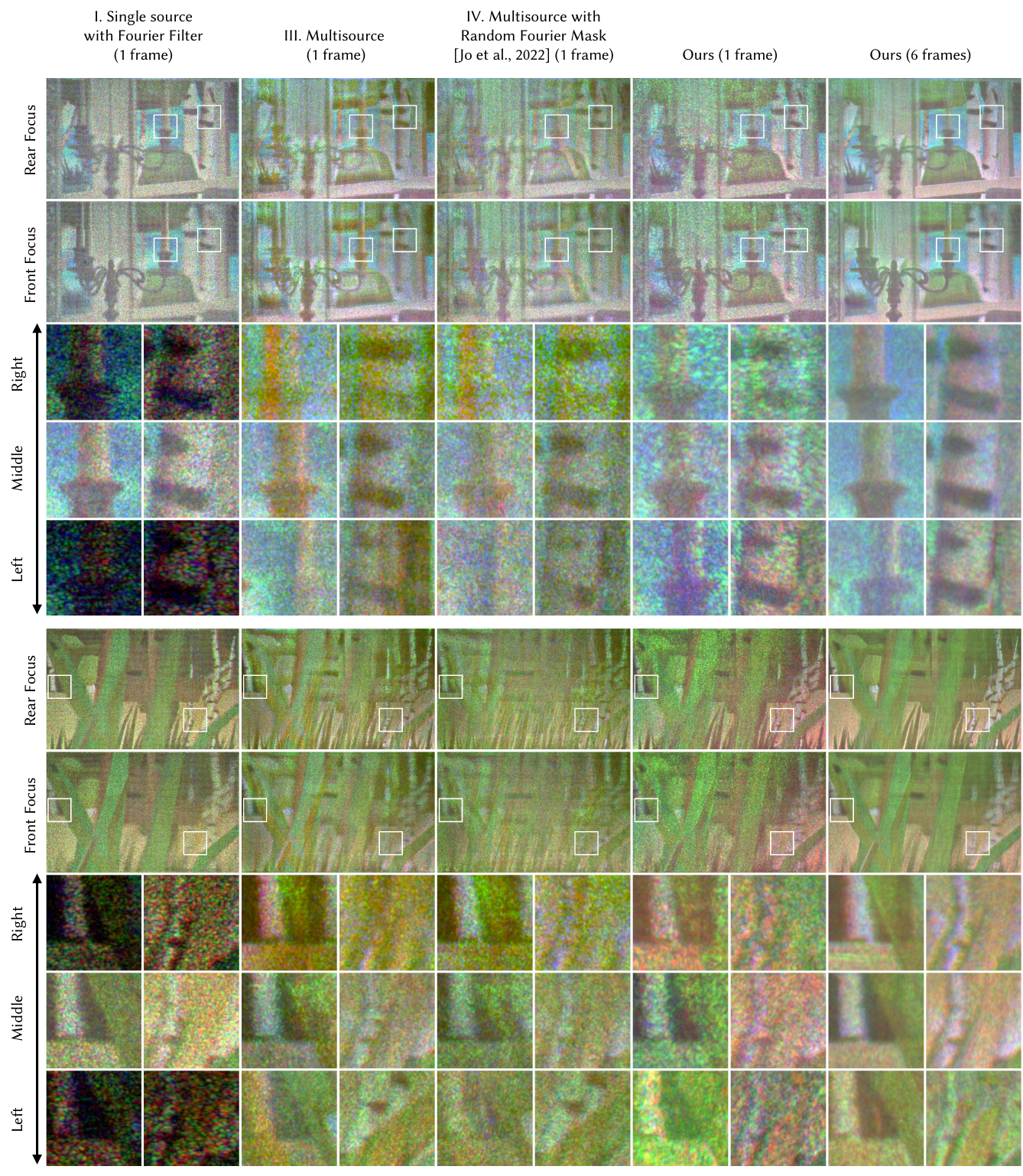}
    \caption{ \textbf{Focals stack and parallax comparison using experimentally captured results.} We compare the conventional setup and various multi-source holographic display configurations, including the single source setup with filter (1st column), the multisource setup without a filter (2nd column), with a random Fourier mask ~\cite{jo2022binary} (3rd column), and our configuration with 1-frame (4th column) and 6-frame time multiplexing (5th column). We show the full image from the central viewpoint in the top row, and the insets in the following rows are captured from different viewpoints. We see that the single-source configuration suffers from extreme brightness falloff at peripheral views. Our method achieves the best image quality among all multisource settings.}
    \label{fig:experimental}
\end{figure*}

\clearpage
\bibliographystyle{ACM-Reference-Format}
\bibliography{references}


\begin{thebibliography}{56}


\ifx \showCODEN    \undefined \def \showCODEN     #1{\unskip}     \fi
\ifx \showDOI      \undefined \def \showDOI       #1{#1}\fi
\ifx \showISBNx    \undefined \def \showISBNx     #1{\unskip}     \fi
\ifx \showISBNxiii \undefined \def \showISBNxiii  #1{\unskip}     \fi
\ifx \showISSN     \undefined \def \showISSN      #1{\unskip}     \fi
\ifx \showLCCN     \undefined \def \showLCCN      #1{\unskip}     \fi
\ifx \shownote     \undefined \def \shownote      #1{#1}          \fi
\ifx \showarticletitle \undefined \def \showarticletitle #1{#1}   \fi
\ifx \showURL      \undefined \def \showURL       {\relax}        \fi
\providecommand\bibfield[2]{#2}
\providecommand\bibinfo[2]{#2}
\providecommand\natexlab[1]{#1}
\providecommand\showeprint[2][]{arXiv:#2}

\bibitem[An et~al\mbox{.}(2020)]%
        {an2020slim}
\bibfield{author}{\bibinfo{person}{Jungkwuen An}, \bibinfo{person}{Kanghee Won}, \bibinfo{person}{Young Kim}, \bibinfo{person}{Jong-Young Hong}, \bibinfo{person}{Hojung Kim}, \bibinfo{person}{Yongkyu Kim}, \bibinfo{person}{Hoon Song}, \bibinfo{person}{Chilsung Choi}, \bibinfo{person}{Yunhee Kim}, \bibinfo{person}{Juwon Seo}, \bibinfo{person}{Alexander Morozov}, \bibinfo{person}{Hyunsik Park}, \bibinfo{person}{Sunghoon Hong}, \bibinfo{person}{Sungwoo Hwang}, \bibinfo{person}{Kichul Kim}, {and} \bibinfo{person}{Hong-Seok Lee}.} \bibinfo{year}{2020}\natexlab{}.
\newblock \showarticletitle{Slim-panel holographic video display}.
\newblock \bibinfo{journal}{\emph{Nature Communications}} \bibinfo{volume}{11}, \bibinfo{number}{1} (\bibinfo{date}{10 Nov} \bibinfo{year}{2020}), \bibinfo{pages}{5568}.
\newblock
\showISSN{2041-1723}
\urldef\tempurl%
\url{https://doi.org/10.1038/s41467-020-19298-4}
\showDOI{\tempurl}


\bibitem[Benton and Bove~Jr(2008)]%
        {benton2008holographic}
\bibfield{author}{\bibinfo{person}{Stephen~A Benton} {and} \bibinfo{person}{V~Michael Bove~Jr}.} \bibinfo{year}{2008}\natexlab{}.
\newblock \bibinfo{booktitle}{\emph{Holographic imaging}}.
\newblock \bibinfo{publisher}{John Wiley \& Sons}.
\newblock


\bibitem[Bigler et~al\mbox{.}(2018)]%
        {bigler2018expansion}
\bibfield{author}{\bibinfo{person}{Colton~M. Bigler}, \bibinfo{person}{Pierre-Alexandre Blanche}, {and} \bibinfo{person}{Kalluri Sarma}.} \bibinfo{year}{2018}\natexlab{}.
\newblock \showarticletitle{Holographic waveguide heads-up display for longitudinal image magnification and pupil expansion}.
\newblock \bibinfo{journal}{\emph{Appl. Opt.}} \bibinfo{volume}{57}, \bibinfo{number}{9} (\bibinfo{date}{Mar} \bibinfo{year}{2018}), \bibinfo{pages}{2007--2013}.
\newblock
\urldef\tempurl%
\url{https://doi.org/10.1364/AO.57.002007}
\showDOI{\tempurl}


\bibitem[Bigler et~al\mbox{.}(2019)]%
        {bigler2019expansion}
\bibfield{author}{\bibinfo{person}{Colton~M. Bigler}, \bibinfo{person}{Micah~S. Mann}, {and} \bibinfo{person}{Pierre-Alexandre Blanche}.} \bibinfo{year}{2019}\natexlab{}.
\newblock \showarticletitle{Holographic waveguide HUD with in-line pupil expansion and 2D FOV expansion}.
\newblock \bibinfo{journal}{\emph{Appl. Opt.}} \bibinfo{volume}{58}, \bibinfo{number}{34} (\bibinfo{date}{Dec} \bibinfo{year}{2019}), \bibinfo{pages}{G326--G331}.
\newblock
\urldef\tempurl%
\url{https://doi.org/10.1364/AO.58.00G326}
\showDOI{\tempurl}


\bibitem[Buckley et~al\mbox{.}(2006)]%
        {buckley2006mask}
\bibfield{author}{\bibinfo{person}{Edward Buckley}, \bibinfo{person}{Adrian Cable}, \bibinfo{person}{Nic Lawrence}, {and} \bibinfo{person}{Tim Wilkinson}.} \bibinfo{year}{2006}\natexlab{}.
\newblock \showarticletitle{Viewing angle enhancement for two- and three-dimensional holographic displays with random superresolution phase masks}.
\newblock \bibinfo{journal}{\emph{Appl. Opt.}} \bibinfo{volume}{45}, \bibinfo{number}{28} (\bibinfo{date}{Oct} \bibinfo{year}{2006}), \bibinfo{pages}{7334--7341}.
\newblock
\urldef\tempurl%
\url{https://doi.org/10.1364/AO.45.007334}
\showDOI{\tempurl}


\bibitem[Chae et~al\mbox{.}(2023)]%
        {Chae2023-ll}
\bibfield{author}{\bibinfo{person}{Minseok Chae}, \bibinfo{person}{Kiseung Bang}, \bibinfo{person}{Dongheon Yoo}, {and} \bibinfo{person}{Yoonchan Jeong}.} \bibinfo{year}{2023}\natexlab{}.
\newblock \showarticletitle{{\'E}tendue Expansion in Holographic Near Eye Displays through Sparse Eye-box Generation Using Lens Array Eyepiece}.
\newblock \bibinfo{journal}{\emph{ACM Trans. Graph.}} \bibinfo{volume}{42}, \bibinfo{number}{4} (\bibinfo{date}{July} \bibinfo{year}{2023}), \bibinfo{pages}{1--13}.
\newblock


\bibitem[Chakravarthula et~al\mbox{.}(2022)]%
        {chakravarthula2022pupil}
\bibfield{author}{\bibinfo{person}{Praneeth Chakravarthula}, \bibinfo{person}{Seung-Hwan Baek}, \bibinfo{person}{Florian Schiffers}, \bibinfo{person}{Ethan Tseng}, \bibinfo{person}{Grace Kuo}, \bibinfo{person}{Andrew Maimone}, \bibinfo{person}{Nathan Matsuda}, \bibinfo{person}{Oliver Cossairt}, \bibinfo{person}{Douglas Lanman}, {and} \bibinfo{person}{Felix Heide}.} \bibinfo{year}{2022}\natexlab{}.
\newblock \showarticletitle{Pupil-Aware Holography}.
\newblock \bibinfo{journal}{\emph{ACM Trans. Graph.}} \bibinfo{volume}{41}, \bibinfo{number}{6}, Article \bibinfo{articleno}{212} (\bibinfo{date}{nov} \bibinfo{year}{2022}), \bibinfo{numpages}{15}~pages.
\newblock
\showISSN{0730-0301}
\urldef\tempurl%
\url{https://doi.org/10.1145/3550454.3555508}
\showDOI{\tempurl}


\bibitem[Chakravarthula et~al\mbox{.}(2020)]%
        {chakravarthula2020learned}
\bibfield{author}{\bibinfo{person}{Praneeth Chakravarthula}, \bibinfo{person}{Ethan Tseng}, \bibinfo{person}{Tarun Srivastava}, \bibinfo{person}{Henry Fuchs}, {and} \bibinfo{person}{Felix Heide}.} \bibinfo{year}{2020}\natexlab{}.
\newblock \showarticletitle{Learned hardware-in-the-loop phase retrieval for holographic near-eye displays}.
\newblock \bibinfo{journal}{\emph{ACM Transactions on Graphics (TOG)}} \bibinfo{volume}{39}, \bibinfo{number}{6} (\bibinfo{year}{2020}), \bibinfo{pages}{1--18}.
\newblock


\bibitem[Chang et~al\mbox{.}(2020)]%
        {Chang2020-ja}
\bibfield{author}{\bibinfo{person}{C Chang}, \bibinfo{person}{K Bang}, \bibinfo{person}{G Wetzstein}, \bibinfo{person}{B Lee}, {and} \bibinfo{person}{L Gao}.} \bibinfo{year}{2020}\natexlab{}.
\newblock \showarticletitle{Toward the next-generation {VR/AR} optics: a review of holographic near-eye displays from a human-centric perspective}.
\newblock \bibinfo{journal}{\emph{Optica}} (\bibinfo{year}{2020}).
\newblock


\bibitem[Chang et~al\mbox{.}(2019)]%
        {chang2019maxweillian}
\bibfield{author}{\bibinfo{person}{Chenliang Chang}, \bibinfo{person}{Wei Cui}, \bibinfo{person}{Jongchan Park}, {and} \bibinfo{person}{Liang Gao}.} \bibinfo{year}{2019}\natexlab{}.
\newblock \showarticletitle{Computational holographic Maxwellian near-eye display with an expanded eyebox}.
\newblock \bibinfo{journal}{\emph{Scientific Reports}} \bibinfo{volume}{9}, \bibinfo{number}{1} (\bibinfo{date}{10 Dec} \bibinfo{year}{2019}), \bibinfo{pages}{18749}.
\newblock
\showISSN{2045-2322}
\urldef\tempurl%
\url{https://doi.org/10.1038/s41598-019-55346-w}
\showDOI{\tempurl}


\bibitem[Chao et~al\mbox{.}(2023)]%
        {Chao2023-wl}
\bibfield{author}{\bibinfo{person}{Brian Chao}, \bibinfo{person}{Manu Gopakumar}, \bibinfo{person}{Suyeon Choi}, {and} \bibinfo{person}{Gordon Wetzstein}.} \bibinfo{year}{2023}\natexlab{}.
\newblock \showarticletitle{High-brightness holographic projection}.
\newblock \bibinfo{journal}{\emph{Opt. Lett.}} \bibinfo{volume}{48}, \bibinfo{number}{15} (\bibinfo{date}{Aug.} \bibinfo{year}{2023}), \bibinfo{pages}{4041--4044}.
\newblock


\bibitem[Choi et~al\mbox{.}(2022)]%
        {choi2022flex}
\bibfield{author}{\bibinfo{person}{Suyeon Choi}, \bibinfo{person}{Manu Gopakumar}, \bibinfo{person}{Yifan Peng}, \bibinfo{person}{Jonghyun Kim}, \bibinfo{person}{Matthew O'Toole}, {and} \bibinfo{person}{Gordon Wetzstein}.} \bibinfo{year}{2022}\natexlab{}.
\newblock \showarticletitle{Time-Multiplexed Neural Holography: A Flexible Framework for Holographic Near-Eye Displays with Fast Heavily-Quantized Spatial Light Modulators}. In \bibinfo{booktitle}{\emph{ACM SIGGRAPH 2022 Conference Proceedings}} (Vancouver, BC, Canada) \emph{(\bibinfo{series}{SIGGRAPH '22})}. \bibinfo{publisher}{Association for Computing Machinery}, \bibinfo{address}{New York, NY, USA}, Article \bibinfo{articleno}{32}, \bibinfo{numpages}{9}~pages.
\newblock
\showISBNx{9781450393379}
\urldef\tempurl%
\url{https://doi.org/10.1145/3528233.3530734}
\showDOI{\tempurl}


\bibitem[Draper et~al\mbox{.}(2019)]%
        {draper2019expansion}
\bibfield{author}{\bibinfo{person}{Craig~T. Draper}, \bibinfo{person}{Colton~M. Bigler}, \bibinfo{person}{Micah~S. Mann}, \bibinfo{person}{Kalluri Sarma}, {and} \bibinfo{person}{Pierre-Alexandre Blanche}.} \bibinfo{year}{2019}\natexlab{}.
\newblock \showarticletitle{Holographic waveguide head-up display with 2-D pupil expansion and longitudinal image magnification}.
\newblock \bibinfo{journal}{\emph{Appl. Opt.}} \bibinfo{volume}{58}, \bibinfo{number}{5} (\bibinfo{date}{Feb} \bibinfo{year}{2019}), \bibinfo{pages}{A251--A257}.
\newblock
\urldef\tempurl%
\url{https://doi.org/10.1364/AO.58.00A251}
\showDOI{\tempurl}


\bibitem[Draper and Blanche(2021)]%
        {draper2021dof}
\bibfield{author}{\bibinfo{person}{Craig~T. Draper} {and} \bibinfo{person}{Pierre-Alexandre Blanche}.} \bibinfo{year}{2021}\natexlab{}.
\newblock \showarticletitle{Examining aberrations due to depth of field in holographic pupil replication waveguide systems}.
\newblock \bibinfo{journal}{\emph{Appl. Opt.}} \bibinfo{volume}{60}, \bibinfo{number}{6} (\bibinfo{date}{Feb} \bibinfo{year}{2021}), \bibinfo{pages}{1653--1659}.
\newblock
\urldef\tempurl%
\url{https://doi.org/10.1364/AO.417756}
\showDOI{\tempurl}


\bibitem[Draper and Blanche(2022)]%
        {draper2022expansion}
\bibfield{author}{\bibinfo{person}{Craig~T. Draper} {and} \bibinfo{person}{Pierre-Alexandre Blanche}.} \bibinfo{year}{2022}\natexlab{}.
\newblock \showarticletitle{Holographic curved waveguide combiner for HUD/AR with 1-D pupil expansion}.
\newblock \bibinfo{journal}{\emph{Opt. Express}} \bibinfo{volume}{30}, \bibinfo{number}{2} (\bibinfo{date}{Jan} \bibinfo{year}{2022}), \bibinfo{pages}{2503--2516}.
\newblock
\urldef\tempurl%
\url{https://doi.org/10.1364/OE.445091}
\showDOI{\tempurl}


\bibitem[Goodman(2005)]%
        {goodman2005fourier}
\bibfield{author}{\bibinfo{person}{Joseph~W Goodman}.} \bibinfo{year}{2005}\natexlab{}.
\newblock \showarticletitle{Introduction to Fourier optics}.
\newblock \bibinfo{journal}{\emph{Introduction to Fourier optics, 3rd ed., by JW Goodman. Englewood, CO: Roberts \& Co. Publishers, 2005}}  \bibinfo{volume}{1} (\bibinfo{year}{2005}).
\newblock


\bibitem[Gopakumar et~al\mbox{.}(2021)]%
        {gopakumar2021unfiltered}
\bibfield{author}{\bibinfo{person}{Manu Gopakumar}, \bibinfo{person}{Jonghyun Kim}, \bibinfo{person}{Suyeon Choi}, \bibinfo{person}{Yifan Peng}, {and} \bibinfo{person}{Gordon Wetzstein}.} \bibinfo{year}{2021}\natexlab{}.
\newblock \showarticletitle{Unfiltered holography: optimizing high diffraction orders without optical filtering for compact holographic displays}.
\newblock \bibinfo{journal}{\emph{Opt. Lett.}} \bibinfo{volume}{46}, \bibinfo{number}{23} (\bibinfo{date}{Dec} \bibinfo{year}{2021}), \bibinfo{pages}{5822--5825}.
\newblock
\urldef\tempurl%
\url{https://doi.org/10.1364/OL.442851}
\showDOI{\tempurl}


\bibitem[Gopakumar et~al\mbox{.}(2024)]%
        {gopakumar2024ar}
\bibfield{author}{\bibinfo{person}{Manu Gopakumar}, \bibinfo{person}{Gun-Yeal Lee}, \bibinfo{person}{Suyeon Choi}, \bibinfo{person}{Brian Chao}, \bibinfo{person}{Yifan Peng}, \bibinfo{person}{Jonghyun Kim}, {and} \bibinfo{person}{Gordon Wetzstein}.} \bibinfo{year}{2024}\natexlab{}.
\newblock \showarticletitle{Full-colour 3D holographic augmented-reality displays with metasurface waveguides}.
\newblock \bibinfo{journal}{\emph{Nature}} (\bibinfo{date}{08 May} \bibinfo{year}{2024}).
\newblock
\showISSN{1476-4687}
\urldef\tempurl%
\url{https://doi.org/10.1038/s41586-024-07386-0}
\showDOI{\tempurl}


\bibitem[Jang et~al\mbox{.}(2024)]%
        {jang2024waveguide}
\bibfield{author}{\bibinfo{person}{Changwon Jang}, \bibinfo{person}{Kiseung Bang}, \bibinfo{person}{Minseok Chae}, \bibinfo{person}{Byoungho Lee}, {and} \bibinfo{person}{Douglas Lanman}.} \bibinfo{year}{2024}\natexlab{}.
\newblock \showarticletitle{Waveguide holography for 3D augmented reality glasses}.
\newblock \bibinfo{journal}{\emph{Nature Communications}} \bibinfo{volume}{15}, \bibinfo{number}{1} (\bibinfo{date}{02 Jan} \bibinfo{year}{2024}), \bibinfo{pages}{66}.
\newblock
\showISSN{2041-1723}
\urldef\tempurl%
\url{https://doi.org/10.1038/s41467-023-44032-1}
\showDOI{\tempurl}


\bibitem[Jang et~al\mbox{.}(2018)]%
        {jang2018hoe}
\bibfield{author}{\bibinfo{person}{Changwon Jang}, \bibinfo{person}{Kiseung Bang}, \bibinfo{person}{Gang Li}, {and} \bibinfo{person}{Byoungho Lee}.} \bibinfo{year}{2018}\natexlab{}.
\newblock \showarticletitle{Holographic Near-Eye Display with Expanded Eye-Box}.
\newblock \bibinfo{journal}{\emph{ACM Trans. Graph.}} \bibinfo{volume}{37}, \bibinfo{number}{6}, Article \bibinfo{articleno}{195} (\bibinfo{date}{dec} \bibinfo{year}{2018}), \bibinfo{numpages}{14}~pages.
\newblock
\showISSN{0730-0301}
\urldef\tempurl%
\url{https://doi.org/10.1145/3272127.3275069}
\showDOI{\tempurl}


\bibitem[Javidi et~al\mbox{.}(2021)]%
        {Javidi2021-vh}
\bibfield{author}{\bibinfo{person}{Bahram Javidi}, \bibinfo{person}{Artur Carnicer}, \bibinfo{person}{Arun Anand}, \bibinfo{person}{George Barbastathis}, \bibinfo{person}{Wen Chen}, \bibinfo{person}{Pietro Ferraro}, \bibinfo{person}{J~W Goodman}, \bibinfo{person}{Ryoichi Horisaki}, \bibinfo{person}{Kedar Khare}, \bibinfo{person}{Malgorzata Kujawinska}, \bibinfo{person}{Rainer~A Leitgeb}, \bibinfo{person}{Pierre Marquet}, \bibinfo{person}{Takanori Nomura}, \bibinfo{person}{Aydogan Ozcan}, \bibinfo{person}{Yongkeun Park}, \bibinfo{person}{Giancarlo Pedrini}, \bibinfo{person}{Pascal Picart}, \bibinfo{person}{Joseph Rosen}, \bibinfo{person}{Genaro Saavedra}, \bibinfo{person}{Natan~T Shaked}, \bibinfo{person}{Adrian Stern}, \bibinfo{person}{Enrique Tajahuerce}, \bibinfo{person}{Lei Tian}, \bibinfo{person}{Gordon Wetzstein}, {and} \bibinfo{person}{Masahiro Yamaguchi}.} \bibinfo{year}{2021}\natexlab{}.
\newblock \showarticletitle{Roadmap on digital holography [Invited]}.
\newblock \bibinfo{journal}{\emph{Opt. Express, OE}} \bibinfo{volume}{29}, \bibinfo{number}{22} (\bibinfo{date}{Oct.} \bibinfo{year}{2021}), \bibinfo{pages}{35078--35118}.
\newblock


\bibitem[Jo et~al\mbox{.}(2022)]%
        {jo2022binary}
\bibfield{author}{\bibinfo{person}{Youngjin Jo}, \bibinfo{person}{Dongheon Yoo}, \bibinfo{person}{Dukho Lee}, \bibinfo{person}{Minkwan Kim}, {and} \bibinfo{person}{Byoungho Lee}.} \bibinfo{year}{2022}\natexlab{}.
\newblock \showarticletitle{Multi-illumination 3D holographic display using a binary mask}.
\newblock \bibinfo{journal}{\emph{Opt. Lett.}} \bibinfo{volume}{47}, \bibinfo{number}{10} (\bibinfo{date}{May} \bibinfo{year}{2022}), \bibinfo{pages}{2482--2485}.
\newblock
\urldef\tempurl%
\url{https://doi.org/10.1364/OL.455348}
\showDOI{\tempurl}


\bibitem[Kavakl{\i} et~al\mbox{.}(2023)]%
        {Kavakli2023-nd}
\bibfield{author}{\bibinfo{person}{Koray Kavakl{\i}}, \bibinfo{person}{Liang Shi}, \bibinfo{person}{Hakan Urey}, \bibinfo{person}{Wojciech Matusik}, {and} \bibinfo{person}{Kaan Ak{\c s}it}.} \bibinfo{year}{2023}\natexlab{}.
\newblock \showarticletitle{Multi-color Holograms Improve Brightness in Holographic Displays}. In \bibinfo{booktitle}{\emph{{SIGGRAPH} Asia 2023 Conference Papers}}. \bibinfo{publisher}{Association for Computing Machinery}, \bibinfo{pages}{1--11}.
\newblock


\bibitem[Kim et~al\mbox{.}(2022)]%
        {Kim2022-cd}
\bibfield{author}{\bibinfo{person}{Jonghyun Kim}, \bibinfo{person}{Manu Gopakumar}, \bibinfo{person}{Suyeon Choi}, \bibinfo{person}{Yifan Peng}, \bibinfo{person}{Ward Lopes}, {and} \bibinfo{person}{Gordon Wetzstein}.} \bibinfo{year}{2022}\natexlab{}.
\newblock \showarticletitle{Holographic Glasses for Virtual Reality}. In \bibinfo{booktitle}{\emph{{ACM} {SIGGRAPH} 2022 Conference Proceedings}} (Vancouver, BC, Canada) \emph{(\bibinfo{series}{SIGGRAPH '22}, \bibinfo{number}{Article 33})}. \bibinfo{publisher}{Association for Computing Machinery}, \bibinfo{address}{New York, NY, USA}, \bibinfo{pages}{1--9}.
\newblock


\bibitem[Kress and Chatterjee(2020)]%
        {kress2020waveguide}
\bibfield{author}{\bibinfo{person}{Bernard~C Kress} {and} \bibinfo{person}{Ishan Chatterjee}.} \bibinfo{year}{2020}\natexlab{}.
\newblock \showarticletitle{Waveguide combiners for mixed reality headsets: a nanophotonics design perspective}.
\newblock \bibinfo{journal}{\emph{Nanophotonics}} \bibinfo{volume}{10}, \bibinfo{number}{1} (\bibinfo{year}{2020}), \bibinfo{pages}{41--74}.
\newblock


\bibitem[Kuo et~al\mbox{.}(2023)]%
        {grace2023multi}
\bibfield{author}{\bibinfo{person}{Grace Kuo}, \bibinfo{person}{Florian Schiffers}, \bibinfo{person}{Douglas Lanman}, \bibinfo{person}{Oliver Cossairt}, {and} \bibinfo{person}{Nathan Matsuda}.} \bibinfo{year}{2023}\natexlab{}.
\newblock \showarticletitle{Multisource Holography}.
\newblock \bibinfo{journal}{\emph{ACM Trans. Graph.}} \bibinfo{volume}{42}, \bibinfo{number}{6}, Article \bibinfo{articleno}{203} (\bibinfo{date}{dec} \bibinfo{year}{2023}), \bibinfo{numpages}{14}~pages.
\newblock
\showISSN{0730-0301}
\urldef\tempurl%
\url{https://doi.org/10.1145/3618380}
\showDOI{\tempurl}


\bibitem[Kuo et~al\mbox{.}(2020)]%
        {kuo2020mask}
\bibfield{author}{\bibinfo{person}{Grace Kuo}, \bibinfo{person}{Laura Waller}, \bibinfo{person}{Ren Ng}, {and} \bibinfo{person}{Andrew Maimone}.} \bibinfo{year}{2020}\natexlab{}.
\newblock \showarticletitle{High Resolution \'{E}tendue Expansion for Holographic Displays}.
\newblock \bibinfo{journal}{\emph{ACM Trans. Graph.}} \bibinfo{volume}{39}, \bibinfo{number}{4}, Article \bibinfo{articleno}{66} (\bibinfo{date}{aug} \bibinfo{year}{2020}), \bibinfo{numpages}{14}~pages.
\newblock
\showISSN{0730-0301}
\urldef\tempurl%
\url{https://doi.org/10.1145/3386569.3392414}
\showDOI{\tempurl}


\bibitem[Lazarev et~al\mbox{.}(2017)]%
        {lazarev2017lcos}
\bibfield{author}{\bibinfo{person}{Grigory Lazarev}, \bibinfo{person}{Stefanie Bonifer}, \bibinfo{person}{Philip Engel}, \bibinfo{person}{Daniel H{\"o}hne}, {and} \bibinfo{person}{Gunther Notni}.} \bibinfo{year}{2017}\natexlab{}.
\newblock \showarticletitle{{High-resolution LCOS microdisplay with sub-kHz frame rate for high performance, high precision 3D sensor}}. In \bibinfo{booktitle}{\emph{Digital Optical Technologies 2017}}, \bibfield{editor}{\bibinfo{person}{Bernard~C. Kress} {and} \bibinfo{person}{Peter Schelkens}} (Eds.), Vol.~\bibinfo{volume}{10335}. International Society for Optics and Photonics, \bibinfo{publisher}{SPIE}, \bibinfo{pages}{103351B}.
\newblock
\urldef\tempurl%
\url{https://doi.org/10.1117/12.2272367}
\showDOI{\tempurl}


\bibitem[Lee et~al\mbox{.}(2020)]%
        {lee2020wide}
\bibfield{author}{\bibinfo{person}{Byounghyo Lee}, \bibinfo{person}{Dongheon Yoo}, \bibinfo{person}{Jinsoo Jeong}, \bibinfo{person}{Seungjae Lee}, \bibinfo{person}{Dukho Lee}, {and} \bibinfo{person}{Byoungho Lee}.} \bibinfo{year}{2020}\natexlab{}.
\newblock \showarticletitle{Wide-angle speckleless DMD holographic display using structured illumination with temporal multiplexing}.
\newblock \bibinfo{journal}{\emph{Opt. Lett.}} \bibinfo{volume}{45}, \bibinfo{number}{8} (\bibinfo{date}{Apr} \bibinfo{year}{2020}), \bibinfo{pages}{2148--2151}.
\newblock
\urldef\tempurl%
\url{https://doi.org/10.1364/OL.390552}
\showDOI{\tempurl}


\bibitem[Lee et~al\mbox{.}(2022)]%
        {lee2022envelope}
\bibfield{author}{\bibinfo{person}{Dukho Lee}, \bibinfo{person}{Kiseung Bang}, \bibinfo{person}{Seung-Woo Nam}, \bibinfo{person}{Byounghyo Lee}, \bibinfo{person}{Dongyeon Kim}, {and} \bibinfo{person}{Byoungho Lee}.} \bibinfo{year}{2022}\natexlab{}.
\newblock \showarticletitle{Expanding energy envelope in holographic display via mutually coherent multi-directional illumination}.
\newblock \bibinfo{journal}{\emph{Scientific Reports}} \bibinfo{volume}{12}, \bibinfo{number}{1} (\bibinfo{date}{22 Apr} \bibinfo{year}{2022}), \bibinfo{pages}{6649}.
\newblock
\showISSN{2045-2322}
\urldef\tempurl%
\url{https://doi.org/10.1038/s41598-022-10355-0}
\showDOI{\tempurl}


\bibitem[Lin et~al\mbox{.}(2018)]%
        {lin2018wedge}
\bibfield{author}{\bibinfo{person}{Wen-Kai Lin}, \bibinfo{person}{Osamu Matoba}, \bibinfo{person}{Bor-Shyh Lin}, {and} \bibinfo{person}{Wei-Chia Su}.} \bibinfo{year}{2018}\natexlab{}.
\newblock \showarticletitle{Astigmatism and deformation correction for a holographic head-mounted display with a wedge-shaped holographic waveguide}.
\newblock \bibinfo{journal}{\emph{Appl. Opt.}} \bibinfo{volume}{57}, \bibinfo{number}{25} (\bibinfo{date}{Sep} \bibinfo{year}{2018}), \bibinfo{pages}{7094--7101}.
\newblock
\urldef\tempurl%
\url{https://doi.org/10.1364/AO.57.007094}
\showDOI{\tempurl}


\bibitem[Lin et~al\mbox{.}(2020)]%
        {lin2020waveguide}
\bibfield{author}{\bibinfo{person}{Wen-Kai Lin}, \bibinfo{person}{Osamu Matoba}, \bibinfo{person}{Bor-Shyh Lin}, {and} \bibinfo{person}{Wei-Chia Su}.} \bibinfo{year}{2020}\natexlab{}.
\newblock \showarticletitle{Astigmatism correction and quality optimization of computer-generated holograms for holographic waveguide displays}.
\newblock \bibinfo{journal}{\emph{Opt. Express}} \bibinfo{volume}{28}, \bibinfo{number}{4} (\bibinfo{date}{Feb} \bibinfo{year}{2020}), \bibinfo{pages}{5519--5527}.
\newblock
\urldef\tempurl%
\url{https://doi.org/10.1364/OE.381193}
\showDOI{\tempurl}


\bibitem[Maimone et~al\mbox{.}(2017)]%
        {maimone2017holo}
\bibfield{author}{\bibinfo{person}{Andrew Maimone}, \bibinfo{person}{Andreas Georgiou}, {and} \bibinfo{person}{Joel~S. Kollin}.} \bibinfo{year}{2017}\natexlab{}.
\newblock \showarticletitle{Holographic Near-Eye Displays for Virtual and Augmented Reality}.
\newblock \bibinfo{journal}{\emph{ACM Trans. Graph.}} \bibinfo{volume}{36}, \bibinfo{number}{4}, Article \bibinfo{articleno}{85} (\bibinfo{date}{jul} \bibinfo{year}{2017}), \bibinfo{numpages}{16}~pages.
\newblock
\showISSN{0730-0301}
\urldef\tempurl%
\url{https://doi.org/10.1145/3072959.3073624}
\showDOI{\tempurl}


\bibitem[Markley et~al\mbox{.}(2023)]%
        {Markley2023-eg}
\bibfield{author}{\bibinfo{person}{Eric Markley}, \bibinfo{person}{Nathan Matsuda}, \bibinfo{person}{Florian Schiffers}, \bibinfo{person}{Oliver Cossairt}, {and} \bibinfo{person}{Grace Kuo}.} \bibinfo{year}{2023}\natexlab{}.
\newblock \showarticletitle{Simultaneous Color Computer Generated Holography}. In \bibinfo{booktitle}{\emph{{SIGGRAPH} Asia 2023 Conference Papers}} (<conf-loc>, <city>Sydney</city>, <state>NSW</state>, <country>Australia</country>, </conf-loc>) \emph{(\bibinfo{series}{SA '23}, \bibinfo{number}{Article 22})}. \bibinfo{publisher}{Association for Computing Machinery}, \bibinfo{address}{New York, NY, USA}, \bibinfo{pages}{1--11}.
\newblock


\bibitem[Monin et~al\mbox{.}(2022a)]%
        {monin2022mask}
\bibfield{author}{\bibinfo{person}{Sagi Monin}, \bibinfo{person}{Aswin~C. Sankaranarayanan}, {and} \bibinfo{person}{Anat Levin}.} \bibinfo{year}{2022}\natexlab{a}.
\newblock \showarticletitle{Analyzing phase masks for wide étendue holographic displays}. In \bibinfo{booktitle}{\emph{2022 IEEE International Conference on Computational Photography (ICCP)}}. \bibinfo{pages}{1--12}.
\newblock
\urldef\tempurl%
\url{https://doi.org/10.1109/ICCP54855.2022.9887757}
\showDOI{\tempurl}


\bibitem[Monin et~al\mbox{.}(2022b)]%
        {monin2022tilt}
\bibfield{author}{\bibinfo{person}{Sagi Monin}, \bibinfo{person}{Aswin~C. Sankaranarayanan}, {and} \bibinfo{person}{Anat Levin}.} \bibinfo{year}{2022}\natexlab{b}.
\newblock \showarticletitle{Exponentially-wide etendue displays using a tilting cascade}. In \bibinfo{booktitle}{\emph{2022 IEEE International Conference on Computational Photography (ICCP)}}. \bibinfo{pages}{1--12}.
\newblock
\urldef\tempurl%
\url{https://doi.org/10.1109/ICCP54855.2022.9887737}
\showDOI{\tempurl}


\bibitem[Padmanaban et~al\mbox{.}(2019)]%
        {padmanaban2019olas}
\bibfield{author}{\bibinfo{person}{N. Padmanaban}, \bibinfo{person}{Y. Peng}, {and} \bibinfo{person}{G. Wetzstein}.} \bibinfo{year}{2019}\natexlab{}.
\newblock \showarticletitle{{Holographic Near-Eye Displays Based on Overlap-Add Stereograms}}.
\newblock \bibinfo{journal}{\emph{ACM Trans. Graph. (SIGGRAPH Asia)}} \bibinfo{number}{6} (\bibinfo{year}{2019}).
\newblock
Issue 38.


\bibitem[Park et~al\mbox{.}(2019)]%
        {park2019photon}
\bibfield{author}{\bibinfo{person}{Jongchan Park}, \bibinfo{person}{KyeoReh Lee}, {and} \bibinfo{person}{YongKeun Park}.} \bibinfo{year}{2019}\natexlab{}.
\newblock \showarticletitle{Ultrathin wide-angle large-area digital 3D holographic display using a non-periodic photon sieve}.
\newblock \bibinfo{journal}{\emph{Nature Communications}} \bibinfo{volume}{10}, \bibinfo{number}{1} (\bibinfo{date}{21 Mar} \bibinfo{year}{2019}), \bibinfo{pages}{1304}.
\newblock
\showISSN{2041-1723}
\urldef\tempurl%
\url{https://doi.org/10.1038/s41467-019-09126-9}
\showDOI{\tempurl}


\bibitem[Park and Kim(2018)]%
        {Park2018-ge}
\bibfield{author}{\bibinfo{person}{Jae-Hyeung Park} {and} \bibinfo{person}{Seong-Bok Kim}.} \bibinfo{year}{2018}\natexlab{}.
\newblock \showarticletitle{Optical see-through holographic near-eye-display with eyebox steering and depth of field control}.
\newblock \bibinfo{journal}{\emph{Opt. Express}} \bibinfo{volume}{26}, \bibinfo{number}{21} (\bibinfo{date}{Oct.} \bibinfo{year}{2018}), \bibinfo{pages}{27076--27088}.
\newblock


\bibitem[Peng et~al\mbox{.}(2020)]%
        {peng2020neural}
\bibfield{author}{\bibinfo{person}{Y. Peng}, \bibinfo{person}{S. Choi}, \bibinfo{person}{N. Padmanaban}, {and} \bibinfo{person}{G. Wetzstein}.} \bibinfo{year}{2020}\natexlab{}.
\newblock \showarticletitle{{Neural Holography with Camera-in-the-loop Training}}.
\newblock \bibinfo{journal}{\emph{ACM Trans. Graph. (SIGGRAPH Asia)}} (\bibinfo{year}{2020}).
\newblock


\bibitem[Pi et~al\mbox{.}(2022)]%
        {Pi2022-mr}
\bibfield{author}{\bibinfo{person}{Dapu Pi}, \bibinfo{person}{Juan Liu}, {and} \bibinfo{person}{Yongtian Wang}.} \bibinfo{year}{2022}\natexlab{}.
\newblock \showarticletitle{Review of computer-generated hologram algorithms for color dynamic holographic three-dimensional display}.
\newblock \bibinfo{journal}{\emph{Light Sci Appl}} \bibinfo{volume}{11}, \bibinfo{number}{1} (\bibinfo{date}{July} \bibinfo{year}{2022}), \bibinfo{pages}{231}.
\newblock


\bibitem[Reichelt et~al\mbox{.}(2010)]%
        {reichelt2010holo}
\bibfield{author}{\bibinfo{person}{Stephan Reichelt}, \bibinfo{person}{Ralf Haussler}, \bibinfo{person}{Norbert Leister}, \bibinfo{person}{Gerald Futterer}, \bibinfo{person}{Hagen Stolle}, {and} \bibinfo{person}{Armin Schwerdtner}.} \bibinfo{year}{2010}\natexlab{}.
\newblock \showarticletitle{Holographic 3-D Displays - Electro-holography within the Grasp of Commercialization}.
\newblock In \bibinfo{booktitle}{\emph{Advances in Lasers and Electro Optics}}, \bibfield{editor}{\bibinfo{person}{Nelson Costa} {and} \bibinfo{person}{Adolfo Cartaxo}} (Eds.). \bibinfo{publisher}{IntechOpen}, \bibinfo{address}{Rijeka}, Chapter~29.
\newblock
\urldef\tempurl%
\url{https://doi.org/10.5772/8650}
\showDOI{\tempurl}


\bibitem[Schiffers et~al\mbox{.}(2023)]%
        {schiffers2023stochastic}
\bibfield{author}{\bibinfo{person}{Florian Schiffers}, \bibinfo{person}{Praneeth Chakravarthula}, \bibinfo{person}{Nathan Matsuda}, \bibinfo{person}{Grace Kuo}, \bibinfo{person}{Ethan Tseng}, \bibinfo{person}{Douglas Lanman}, \bibinfo{person}{Felix Heide}, {and} \bibinfo{person}{Oliver Cossairt}.} \bibinfo{year}{2023}\natexlab{}.
\newblock \showarticletitle{Stochastic Light Field Holography}.
\newblock \bibinfo{journal}{\emph{IEEE International Conference on Computational Photography (ICCP)}} (\bibinfo{year}{2023}).
\newblock


\bibitem[Shi et~al\mbox{.}(2021)]%
        {shi2021neural}
\bibfield{author}{\bibinfo{person}{Liang Shi}, \bibinfo{person}{Beichen Li}, \bibinfo{person}{Changil Kim}, \bibinfo{person}{Petr Kellnhofer}, {and} \bibinfo{person}{Wojciech Matusik}.} \bibinfo{year}{2021}\natexlab{}.
\newblock \showarticletitle{Towards real-time photorealistic 3D holography with deep neural networks}.
\newblock \bibinfo{journal}{\emph{Nature}} \bibinfo{volume}{591}, \bibinfo{number}{7849} (\bibinfo{date}{01 Mar} \bibinfo{year}{2021}), \bibinfo{pages}{234--239}.
\newblock
\showISSN{1476-4687}
\urldef\tempurl%
\url{https://doi.org/10.1038/s41586-020-03152-0}
\showDOI{\tempurl}


\bibitem[Shi et~al\mbox{.}(2022)]%
        {Shi2022-ih}
\bibfield{author}{\bibinfo{person}{Liang Shi}, \bibinfo{person}{Beichen Li}, {and} \bibinfo{person}{Wojciech Matusik}.} \bibinfo{year}{2022}\natexlab{}.
\newblock \showarticletitle{End-to-end learning of {3D} phase-only holograms for holographic display}.
\newblock \bibinfo{journal}{\emph{Light Sci Appl}} \bibinfo{volume}{11}, \bibinfo{number}{1} (\bibinfo{date}{Aug.} \bibinfo{year}{2022}), \bibinfo{pages}{247}.
\newblock


\bibitem[Shi et~al\mbox{.}(2024)]%
        {shi2024ergo}
\bibfield{author}{\bibinfo{person}{Liang Shi}, \bibinfo{person}{DongHun Ryu}, {and} \bibinfo{person}{Wojciech Matusik}.} \bibinfo{year}{2024}\natexlab{}.
\newblock \showarticletitle{Ergonomic-Centric Holography: Optimizing Realism, Immersion, and Comfort for Holographic Display}.
\newblock \bibinfo{journal}{\emph{Laser \& Photonics Reviews}} \bibinfo{volume}{18}, \bibinfo{number}{4} (\bibinfo{year}{2024}), \bibinfo{pages}{2300651}.
\newblock
\urldef\tempurl%
\url{https://doi.org/10.1002/lpor.202300651}
\showDOI{\tempurl}


\bibitem[Starikov(1982)]%
        {starikov1982effective}
\bibfield{author}{\bibinfo{person}{A Starikov}.} \bibinfo{year}{1982}\natexlab{}.
\newblock \showarticletitle{Effective number of degrees of freedom of partially coherent sources}.
\newblock \bibinfo{journal}{\emph{JOSA}} \bibinfo{volume}{72}, \bibinfo{number}{11} (\bibinfo{year}{1982}), \bibinfo{pages}{1538--1544}.
\newblock


\bibitem[Sun et~al\mbox{.}(2013)]%
        {sun2013large}
\bibfield{author}{\bibinfo{person}{Jie Sun}, \bibinfo{person}{Erman Timurdogan}, \bibinfo{person}{Ami Yaacobi}, \bibinfo{person}{Ehsan~Shah Hosseini}, {and} \bibinfo{person}{Michael~R Watts}.} \bibinfo{year}{2013}\natexlab{}.
\newblock \showarticletitle{Large-scale nanophotonic phased array}.
\newblock \bibinfo{journal}{\emph{Nature}} \bibinfo{volume}{493}, \bibinfo{number}{7431} (\bibinfo{year}{2013}), \bibinfo{pages}{195--199}.
\newblock


\bibitem[Tseng et~al\mbox{.}(2024)]%
        {tseng2023neural}
\bibfield{author}{\bibinfo{person}{Ethan Tseng}, \bibinfo{person}{Grace Kuo}, \bibinfo{person}{Seung-Hwan Baek}, \bibinfo{person}{Nathan Matsuda}, \bibinfo{person}{Andrew Maimone}, \bibinfo{person}{Florian Schiffers}, \bibinfo{person}{Praneeth Chakravarthula}, \bibinfo{person}{Qiang Fu}, \bibinfo{person}{Wolfgang Heidrich}, \bibinfo{person}{Douglas Lanman}, {and} \bibinfo{person}{Felix Heide}.} \bibinfo{year}{2024}\natexlab{}.
\newblock \showarticletitle{Neural {\'e}tendue expander for ultra-wide-angle high-fidelity holographic display}.
\newblock \bibinfo{journal}{\emph{Nature Communications}} \bibinfo{volume}{15}, \bibinfo{number}{1} (\bibinfo{date}{22 Apr} \bibinfo{year}{2024}), \bibinfo{pages}{2907}.
\newblock
\showISSN{2041-1723}
\urldef\tempurl%
\url{https://doi.org/10.1038/s41467-024-46915-3}
\showDOI{\tempurl}


\bibitem[Wang et~al\mbox{.}(2023)]%
        {Wang2023-vf}
\bibfield{author}{\bibinfo{person}{Zi Wang}, \bibinfo{person}{Guoqiang Lv}, \bibinfo{person}{Yujian Pang}, \bibinfo{person}{Qibin Feng}, \bibinfo{person}{Anting Wang}, {and} \bibinfo{person}{Hai Ming}.} \bibinfo{year}{2023}\natexlab{}.
\newblock \showarticletitle{Lens array-based holographic {3D} display with an expanded field of view and eyebox}.
\newblock \bibinfo{journal}{\emph{Opt. Lett.}} \bibinfo{volume}{48}, \bibinfo{number}{21} (\bibinfo{date}{Nov.} \bibinfo{year}{2023}), \bibinfo{pages}{5559--5562}.
\newblock


\bibitem[Xia et~al\mbox{.}(2020)]%
        {Xia2020-sa}
\bibfield{author}{\bibinfo{person}{Xinxing Xia}, \bibinfo{person}{Yunqing Guan}, \bibinfo{person}{{Andrei State}}, \bibinfo{person}{Praneeth Chakravarthula}, \bibinfo{person}{Tat-Jen Cham}, {and} \bibinfo{person}{Henry Fuchs}.} \bibinfo{year}{2020}\natexlab{}.
\newblock \showarticletitle{Towards Eyeglass-style Holographic Near-eye Displays with Statically}. In \bibinfo{booktitle}{\emph{2020 {IEEE} International Symposium on Mixed and Augmented Reality ({ISMAR})}}. \bibinfo{publisher}{IEEE}, \bibinfo{pages}{312--319}.
\newblock


\bibitem[Xia et~al\mbox{.}(2023)]%
        {xia2023steer}
\bibfield{author}{\bibinfo{person}{Xinxing Xia}, \bibinfo{person}{Weisen Wang}, \bibinfo{person}{Frank Guan}, \bibinfo{person}{Furong Yang}, \bibinfo{person}{Xinghua Shui}, \bibinfo{person}{Huadong Zheng}, \bibinfo{person}{Yingjie Yu}, {and} \bibinfo{person}{Yifan Peng}.} \bibinfo{year}{2023}\natexlab{}.
\newblock \showarticletitle{Exploring angular-steering illumination-based eyebox expansion for holographic displays}.
\newblock \bibinfo{journal}{\emph{Opt. Express}} \bibinfo{volume}{31}, \bibinfo{number}{19} (\bibinfo{date}{Sep} \bibinfo{year}{2023}), \bibinfo{pages}{31563--31573}.
\newblock
\urldef\tempurl%
\url{https://opg.optica.org/oe/abstract.cfm?URI=oe-31-19-31563}
\showURL{%
\tempurl}


\bibitem[Yang et~al\mbox{.}(2022)]%
        {yang2022diffeng}
\bibfield{author}{\bibinfo{person}{Daeho Yang}, \bibinfo{person}{Wontaek Seo}, \bibinfo{person}{Hyeonseung Yu}, \bibinfo{person}{Sun~Il Kim}, \bibinfo{person}{Bongsu Shin}, \bibinfo{person}{Chang-Kun Lee}, \bibinfo{person}{Seokil Moon}, \bibinfo{person}{Jungkwuen An}, \bibinfo{person}{Jong-Young Hong}, \bibinfo{person}{Geeyoung Sung}, {and} \bibinfo{person}{Hong-Seok Lee}.} \bibinfo{year}{2022}\natexlab{}.
\newblock \showarticletitle{Diffraction-engineered holography: Beyond the depth representation limit of holographic displays}.
\newblock \bibinfo{journal}{\emph{Nature Communications}} \bibinfo{volume}{13}, \bibinfo{number}{1} (\bibinfo{date}{12 Oct} \bibinfo{year}{2022}), \bibinfo{pages}{6012}.
\newblock
\showISSN{2041-1723}
\urldef\tempurl%
\url{https://doi.org/10.1038/s41467-022-33728-5}
\showDOI{\tempurl}


\bibitem[Yeom et~al\mbox{.}(2021)]%
        {yeom2021voxel}
\bibfield{author}{\bibinfo{person}{Jiwoon Yeom}, \bibinfo{person}{Yeseul Son}, {and} \bibinfo{person}{Kwangsoon Choi}.} \bibinfo{year}{2021}\natexlab{}.
\newblock \showarticletitle{Crosstalk Reduction in Voxels for a See-Through Holographic Waveguide by Using Integral Imaging with Compensated Elemental Images}.
\newblock \bibinfo{journal}{\emph{Photonics}} \bibinfo{volume}{8}, \bibinfo{number}{6} (\bibinfo{year}{2021}).
\newblock
\showISSN{2304-6732}
\urldef\tempurl%
\url{https://doi.org/10.3390/photonics8060217}
\showDOI{\tempurl}


\bibitem[Yu et~al\mbox{.}(2017)]%
        {yu2017speckle}
\bibfield{author}{\bibinfo{person}{Hyeonseung Yu}, \bibinfo{person}{KyeoReh Lee}, \bibinfo{person}{Jongchan Park}, {and} \bibinfo{person}{YongKeun Park}.} \bibinfo{year}{2017}\natexlab{}.
\newblock \showarticletitle{Ultrahigh-definition dynamic 3D holographic display by active control of volume speckle fields}.
\newblock \bibinfo{journal}{\emph{Nature Photonics}} \bibinfo{volume}{11}, \bibinfo{number}{3} (\bibinfo{date}{01 Mar} \bibinfo{year}{2017}), \bibinfo{pages}{186--192}.
\newblock
\showISSN{1749-4893}
\urldef\tempurl%
\url{https://doi.org/10.1038/nphoton.2016.272}
\showDOI{\tempurl}


\bibitem[Zhang and Levoy(2009)]%
        {zhang2009wigner}
\bibfield{author}{\bibinfo{person}{Zhengyun Zhang} {and} \bibinfo{person}{Marc Levoy}.} \bibinfo{year}{2009}\natexlab{}.
\newblock \showarticletitle{Wigner distributions and how they relate to the light field}. In \bibinfo{booktitle}{\emph{2009 IEEE International Conference on Computational Photography (ICCP)}}. IEEE, \bibinfo{pages}{1--10}.
\newblock


\end{thebibliography}

\end{document}